\documentclass[a4paper,11pt]{article}
\usepackage{jinstpub} 
\usepackage{lineno}

\usepackage{multirow}
 \usepackage{booktabs}

\title{\boldmath Response of AC-coupled Low Gain Avalanche Detectors to Ionizing and Non-ionizing Radiation Damage}

\author[a]{J. Si,}
\author[b]{G. D'Amen,}
\author[a]{M. H. Mohamed Farook,}
\author[b]{G. Giacomini,}
\author[a]{M. R. Hoeferkamp,}
\author[a]{S.~Seidel,}
\author[b]{A. Tricoli}
\affiliation[a]{Department of Physics and Astronomy, University of New Mexico, Albuquerque, NM 87131}
\affiliation[b]{Physics Department, Brookhaven National Laboratory, Upton, NY 11973}

\emailAdd{seidel@unm.edu}

\abstract{Low gain avalanche diodes with DC- and AC-coupled readout were exposed to ionizing and non-ionizing radiation at levels relevant to future experiments in particle, nuclear, and medical physics and to astrophysics.  Damage-related change in their acceptor removal constants and in the resistivity of the region between the guard ring and the active area are reported, as is change in the leakage current and depletion voltages of the active volumes.}


\arxivnumber{2508.21028} 

\begin{document}
\maketitle
\flushbottom

\section{Introduction}
\label{sec:intro}
Low gain avalanche diodes (LGADs) of a design in which the gain layer is not segmented and the signal is read out via capacitive coupling have been studied.  These AC-LGADs were designed and fabricated at Brookhaven National Laboratory (BNL).  They were then irradiated with protons and gammas and evaluated at the University of New Mexico.  A similar production of single-channel LGADs (DC-LGADs) was subjected to a similar analysis to allow comparisons between technologies.

\begin{figure}[htbp]
\centering
\includegraphics[width=.25\textwidth]{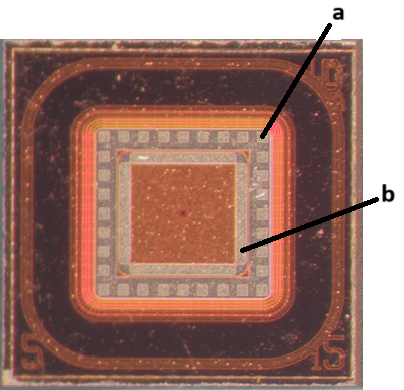}
\qquad
\includegraphics[width=.25\textwidth]{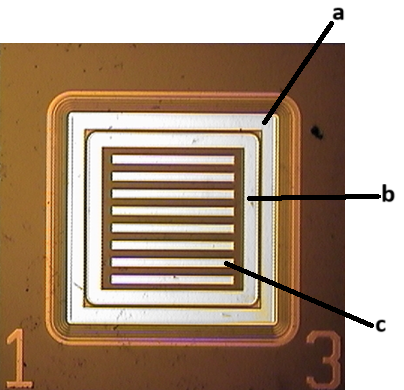}
\qquad
\includegraphics[width=.25\textwidth]{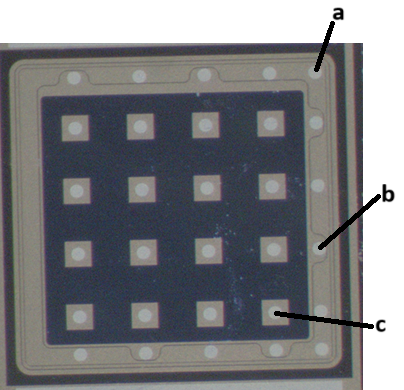}
\caption{DC-LGAD from wafer 3076 (left); AC-LGAD from wafer 3073 (center); and AC-LGAD from wafer 3080 (right).  In each figure, the labels indicate (a) the guard ring, (b) the DC contact pad, and (c) the AC pad or strip, which appears on the AC-LGAD devices only.  \label{fig:photos}}
\end{figure}

The LGAD technology was invented to provide high precision in measurements of particle time of detection~\cite{pellegrini, cartiglia, sadrozinski}.
In the DC-LGAD design, junction termination extensions (JTE) and p-stops or p-spray are needed to control the electric field between channels.  For DC-LGADs, the pad dimension must be larger than the substrate thickness to guarantee a uniform electric field, and hence uniform gain, in the gain layer.
The capacitive coupling of the AC-LGAD~\cite{giacomini, mandurrino, tornago}
eliminates need for most of the field-shaping features; the AC-LGAD configuration thus permits high fill factor across the full device.  The continuous gain layer of the AC-LGAD design can permit a smaller pitch than is possible in the DC case.  Studies have shown some LGAD configurations to maintain good timing resolution under non-ionizing radiation on the order of about $2.5 \times 10^{15}$ n$_{\rm{eq}}/{\rm{cm}^2}$~\cite{ferrero} 
and ionizing radiation approaching 2 MGy~\cite{hoeferkamp}.
While these tolerances are suitable~\cite{atlastdr, cms}
for expected conditions at middle radii at the HL-LHC~\cite{hllhc} experiments up to integrated luminosity of 4000~fb$^{-1}$, higher tolerance is sought for future applications and longer durations.  Table~\ref{tab:i} summarizes the features of the LGADs that were studied.  
Figure~\ref{fig:photos} shows photographs of example devices that were studied.

\begin{table}[htbp]
\centering
\caption{Features of the devices studied.\label{tab:i}}
\smallskip
\begin{tabular}{c|c|c|c}
\hline
&AC-LGADs&DC-LGADs&AC-LGADs\\
&wafer 3073&wafer 3076&wafer 3080\\
\hline
Configuration&strip&pixel&pixel\\
Area of the n$^{++}$ layer (mm$^2$)&$2.05 \times 2.05$&$1.3 \times 1.3$&$2.05 \times 2.05$\\
Area of the gain layer (mm$^2)$&$1.95 \times 1.95$&$1.2 \times 1.2$&$1.95 \times 1.95$\\
Thickness of the active volume ($\mu$m)&20&20&20\\
Applied radiation species&gamma&proton&proton\\

\hline
\end{tabular}
\end{table}

The devices were separated into several sample groups: some were retained unirradiated as controls; some were exposed at the Sandia National Laboratories Gamma Irradiation Facility~\cite{snl} to gamma doses from 0.4 MGy to 10 MGy; and some were exposed to 400-MeV protons at the Fermilab Irradiation Test Area~\cite{fnalita} to fluences 
up to $2.06 \times 10^{15}$ n$_{\rm eq}/{\rm cm}^2$.  A hardness factor of 0.83 was used~\cite{niel} to convert the proton damage to the equivalent 1-MeV neutron damage.  The gamma doses are known to 5\% uncertainty, and the uncertainties on the proton fluences range from 10\% to 11\%.  Subsequent to the proton irradiation, devices were annealed at $60^\circ$C for 80 minutes.  Proton- and gamma-irradiated devices were held in the freezer below $-28^\circ$C when they were not being annealed or measured.

\section{Gamma Irradiation Studies}
\label{sect:gamma}

\begin{figure}[htbp]
\centering
\includegraphics[width=.6\textwidth]{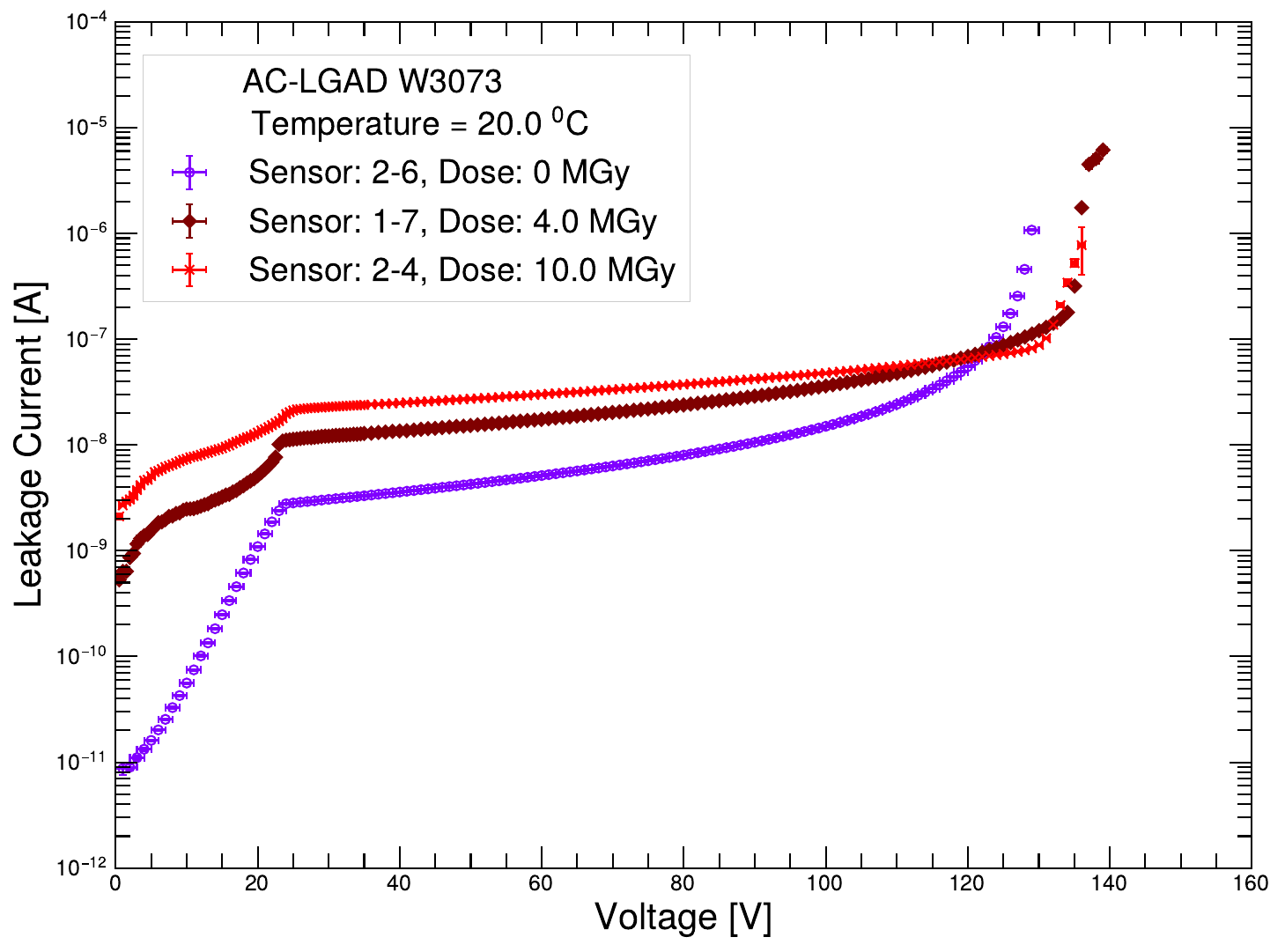}
\caption{The leakage current versus bias voltage of AC-LGAD strip sensors for three levels of gamma exposure. (The uncertainty bars on data points below the depletion voltage have been excluded for improved visualization.)  The data were collected at temperatures in the range $19.8^\circ$C to $20.0^\circ$C and are all scaled to $20^\circ$C for ease of comparison.
\label{fig:W3073_IV}}
\end{figure}

Figure~\ref{fig:W3073_IV} shows the leakage current per channel (excluding guard ring current) of the strip-geometry AC-LGADs (Wafer 3073), as a function of bias voltage ("IV"), for three levels of gamma exposure.  The channel current rises from about 5~nA unirradiated to about 30 nA at 10 MGy. The breakdown voltage increases from about 120 V unirradiated to about 130 V at 10 MGy.
To characterize the effects of the irradiation upon the leakage current of the samples shown in Figure~\ref{fig:W3073_IV}, the data, which were collected at different temperatures, are all scaled to a reference temperature of $20^\circ$C through the relation~\cite{Chilingarov:2013adb}:

\begin{equation}
    I_{\text{leak}}(T_{\text{ref}}) = I_{\text{leak}}(T) \cdot \left(\frac{T_{\text{ref}}}{T}\right)^2 \cdot e^{-\frac{E_{\text{eff}}}{k_B} \left(\frac{1}{T_{\text{ref}}} - \frac{1}{T}\right)}.
    \label{eq:temp_scaling_leakage}
\end{equation}
Here, $T_{\text{ref}}$ is the reference temperature (in Kelvin) to which the leakage current $I_{\text{leak}}$ is scaled, $T$ is the measured temperature (in Kelvin) of the sensor, $k_B$ is the Boltzmann constant, and $E_{\text{eff}}$ is the effective band gap of the (irradiated) sensor, 1.21~eV. The scaled data are shown in Table~\ref{tab:leakage_current_W3073}.

\begin{table}[htbp]
\centering
\caption{Leakage current of AC-LGADs (W3073), scaled to $20^\circ$ C, measured at 50~V for several gamma doses.  Because pre-irradiation leakage current data are not available for Sensors 1-7 and 2-4, their post-irradiation values are compared to the unirradiated current value of Sensor 2-6. $\mathcal R$ is the ratio of their post-irradiation currents to the current of the unirradiated Sensor 2-6 which has the same design.  Sensor 1-3 had some irregularities in its guard ring which prevented reliable leakage current measurement but did not prevent measurements of capacitance.  The uncertainties on these central values are discussed in Appendix~\ref{app:B}.}
\label{tab:leakage_current_W3073}
\begin{tabular}{ccc c}
\toprule
{Dose} & {Sensor} & {$I_{\rm leakage}$} & {$\mathcal R$} \\

[MGy]&number&[nA]\\

\midrule
0     & 2--6 & 4.4  & -  \\
0.4&1--3&-&-\\
4.0    & 1--7 & 15.3 & 3.5 $\pm$ 0.1 \\
10.0  & 2--4 & 27.3 & 6.2 $\pm$ 0.2 \\
\bottomrule
\end{tabular}
\end{table}

\begin{figure}[htbp]
\centering
\includegraphics[width=.45\textwidth]{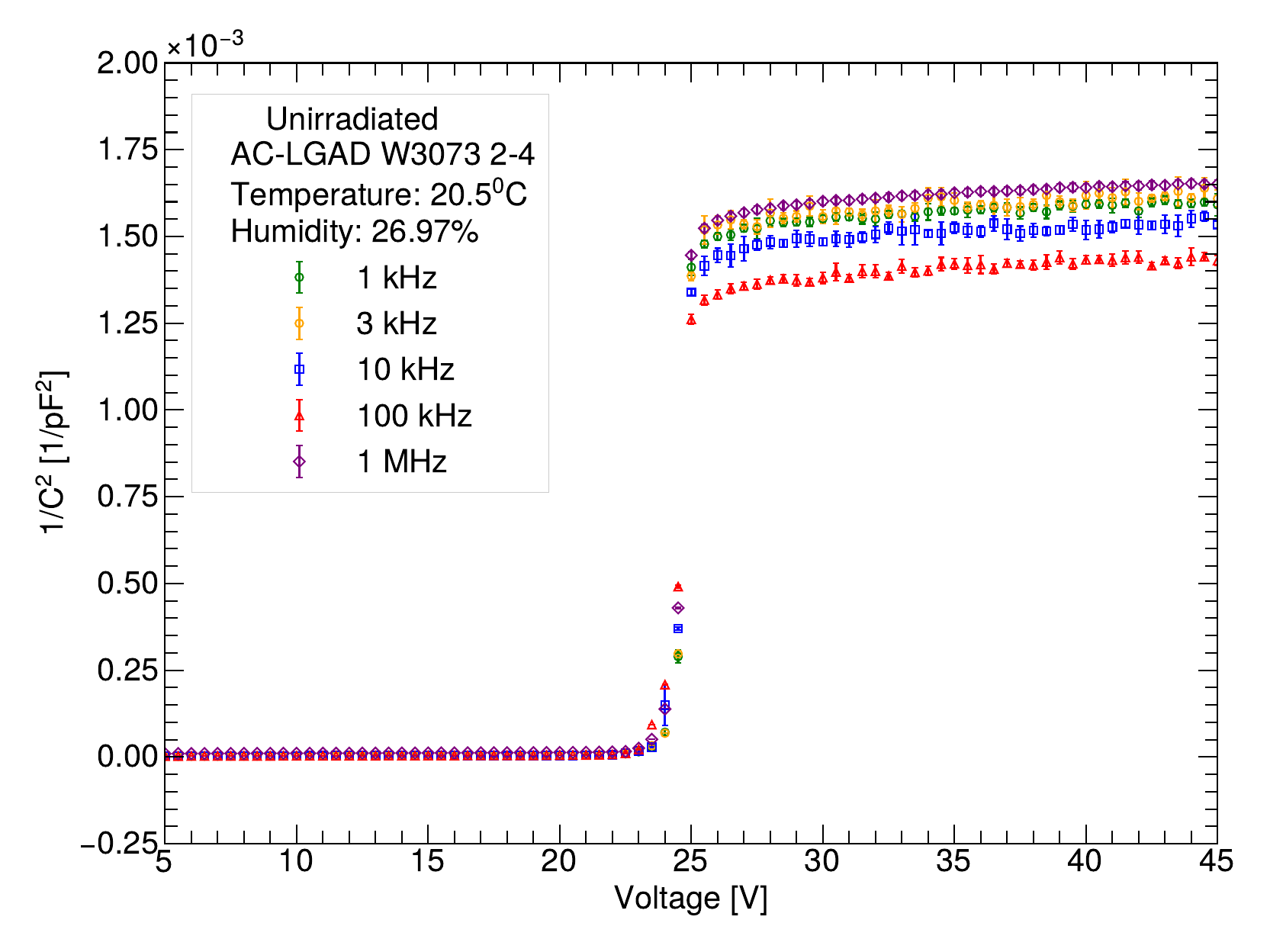}
\qquad
\includegraphics[width=.45\textwidth]{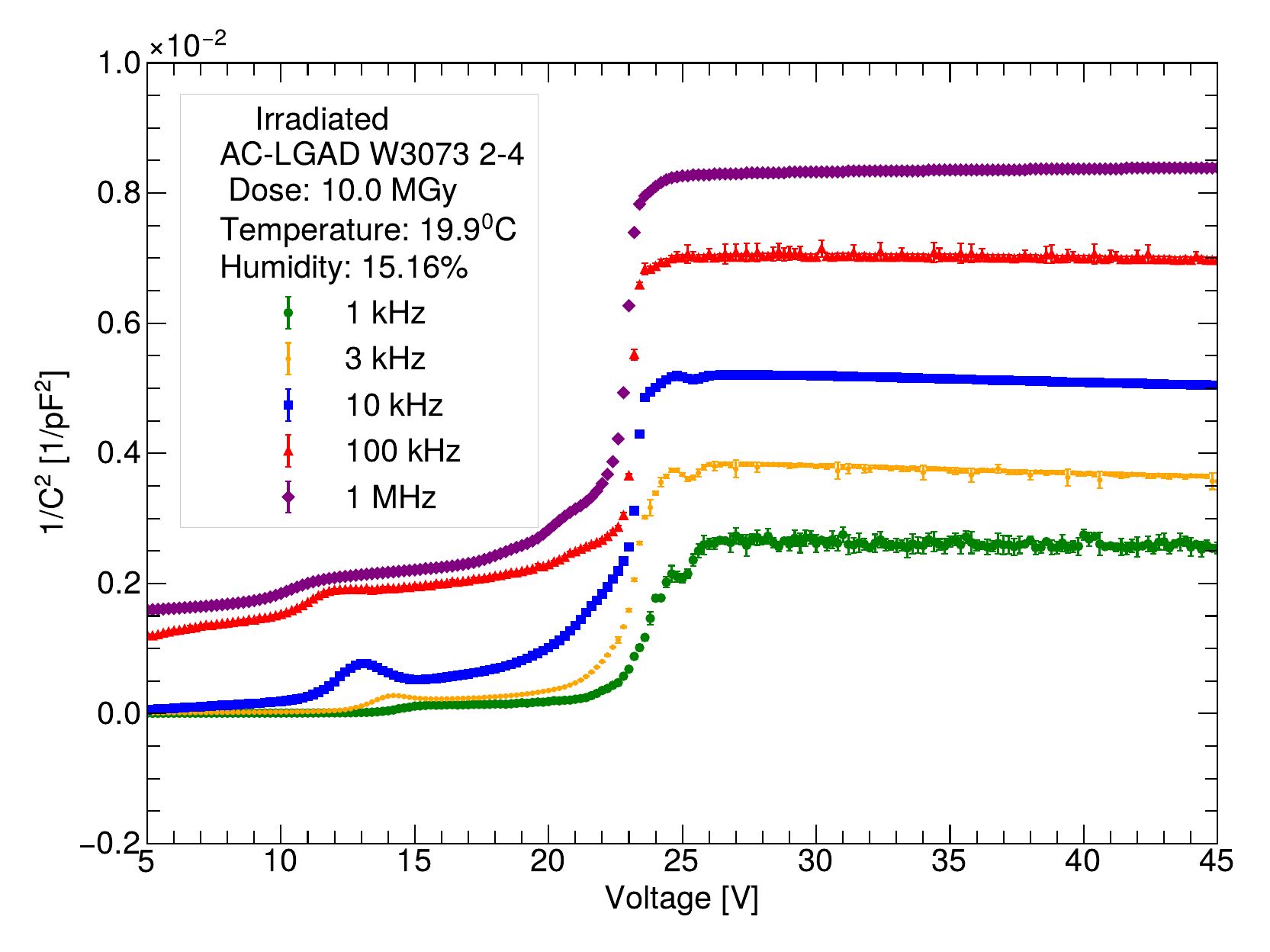}
\caption{Inverse capacitance (C$^{-2}$) versus bias voltage of AC-LGAD strip sensors for two levels of gamma exposure and five applied signal frequencies.  The unirradiated case is shown on the left, and the irradiated, on the right. The legends show the temperatures at which the data were collected.\label{fig:W3073_CV}}
\end{figure}

Figure~\ref{fig:W3073_CV} shows the inverse capacitance C$^{-2}$ versus bias voltage ("CV") for the strip-geometry AC-LGADs, for gamma exposures of zero and 10 MGy and five applied signal frequencies. While the gamma irradiation process evidently produces frequency-dependence of the observed capacitance values, all five applied frequencies yield similar results on the depletion voltages to be extracted. Data collected at signal frequency 10 kHz are selected to extract the gain layer depletion voltage V$_{\rm gl}$ and the full depletion voltage V$_{\rm fd}$.  The method to extract those voltages is illustrated in Appendix~\ref{app:depletionVoltage}. First, the C$^{-2}$ versus V trend is smoothed using a Savitzky-Golay filter function~\cite{Savitzky:2002vxy} that applies local polynomial regression on subsets of adjacent data points to smooth noisy data while preserving the shape and features of the underlying signal.  Boundaries  separating three regions in the trend are obtained by taking the second derivative of the smoothed function and noting the two points $V_1$ and $V_2$ at which it changes significantly.  Three lines are fitted to the three regions of the trend, and the V$_{\rm gl}$ and V$_{\rm fd}$ values are extracted from the lower and upper intersection points of those lines, respectively.  Table~\ref{tab:vgl_vfd_gamma_irr} summarizes the measured values of gain layer depletion voltage, full depletion voltage, and bulk depletion voltage ${\rm V}_{\rm bulk}$, which is defined as ${\rm V}_{\rm fd}-{\rm V}_{\rm gl}$, all before and after gamma irradiation.  The method for determining the uncertainties on leakage current, capacitance, and depletion voltages is described in Appendix~\ref{app:B}.  The uncertainties on ${\rm V}_{\rm gl}$ and ${\rm V}_{\rm fd}$ include uncertainties associated with the fitting process.  Application of this level of gamma dose appears to have negligible effect on depletion voltages, possibly reducing them by as much as one volt, but this is within the margin of uncertainty on the measurement.




\begin{table}[htbp]
\centering
\caption{Sensor voltages before and after gamma irradiation to various doses.}
\label{tab:vgl_vfd_gamma_irr}
\small
\begin{tabular}{c|ccc|c|ccc}
\toprule
\multirow{2}{*}{\textbf{Sensor}} & \multicolumn{3}{c|}{\textbf{Unirradiated}} & \multirow{2}{*}{\begin{tabular}[c]{@{}c@{}}\textbf{Dose} \\ {[MGy]}\end{tabular}} & \multicolumn{3}{c}{\textbf{Irradiated}} \\ \cmidrule{2-4} \cmidrule{6-8} & $V_{\rm gl}$ [V] & $V_{\rm fd}$ [V] & $V_{\rm bulk}$ [V] & & $V_{\rm gl}$ [V] & $V_{\rm fd}$ [V] & $V_{\rm bulk}$ [V] \\ \midrule
2--6 & 23.3 $\pm$ 0.6 & 25.2 $\pm$ 0.6 & 1.9 $\pm$ 0.9 & 0.0 & -- & -- & -- \\
1--3 & 23.3 $\pm$ 0.6 & 25.1 $\pm$ 0.6 & 1.8 $\pm$ 0.8 & 0.4 & 23.2 $\pm$ 0.5 & 24.8 $\pm$ 0.6 & 1.6 $\pm$ 0.8 \\
1--7 & 23.2 $\pm$ 0.6 & 25.1 $\pm$ 0.6 & 1.9 $\pm$ 0.8 & 4.0 & 22.3 $\pm$ 0.5 & 23.8 $\pm$ 0.6 & 1.5 $\pm$ 0.8 \\
2--4 & 23.7 $\pm$ 0.6 & 25.7 $\pm$ 0.6 & 2.0 $\pm$ 0.9 & 10.0 & 21.5 $\pm$ 0.7 & 24.2 $\pm$ 0.4 & 2.7 $\pm$ 0.9 \\ \bottomrule
\end{tabular}
\end{table}

The resistance between the single DC contact and the guard ring was also measured as a function of gamma dose to evaluate the effects of surface damage due to ionizing radiation. The integrity of the region between the guard ring and the DC pad is critical to minimizing leakage current, which contributes to noise.

\begin{figure}[htbp]
\centering
\includegraphics[width=.6\textwidth]{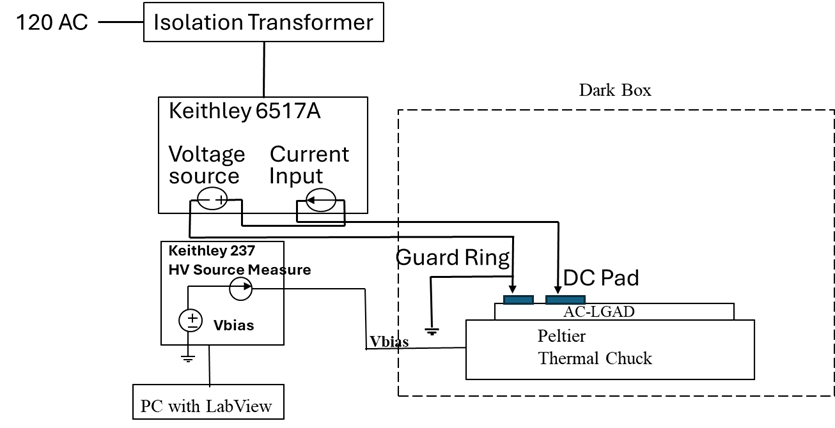}
\caption{Experimental setup for measuring the resistance between the guard ring and readout pads of the LGADs.\label{fig:resistance}}
\end{figure}

The setup for making the measurement is shown in Figure~\ref{fig:resistance}.   To measure the inter-pad resistance, a Keithley 237 source measure unit is used to apply negative bias voltage to the back of the AC-LGAD.  The guard ring of the LGAD is connected to ground.  A Keithley 6517A electrometer is used to apply voltage between the guard ring and the DC pad using its internal voltage source, and the inter-pad current is read by its internal ammeter.  An isolation transformer is used to ensure that the chassis of the Keithley 6517A floats with respect to the input power earth ground.  This eliminates the possibility that any part of the inter-pad current bypasses the internal ammeter and flows to ground through the chassis.  The inter-pad resistance $R_{\rm int}$ is calculated from $R_{\rm int}=V_{\rm int}/I_{\rm int}$, where $V_{\rm int}$ is the inter-pad voltage and $I_{\rm int}$ is the inter-pad current.

\begin{figure}[htbp]
\centering
\includegraphics[width=.6\textwidth]{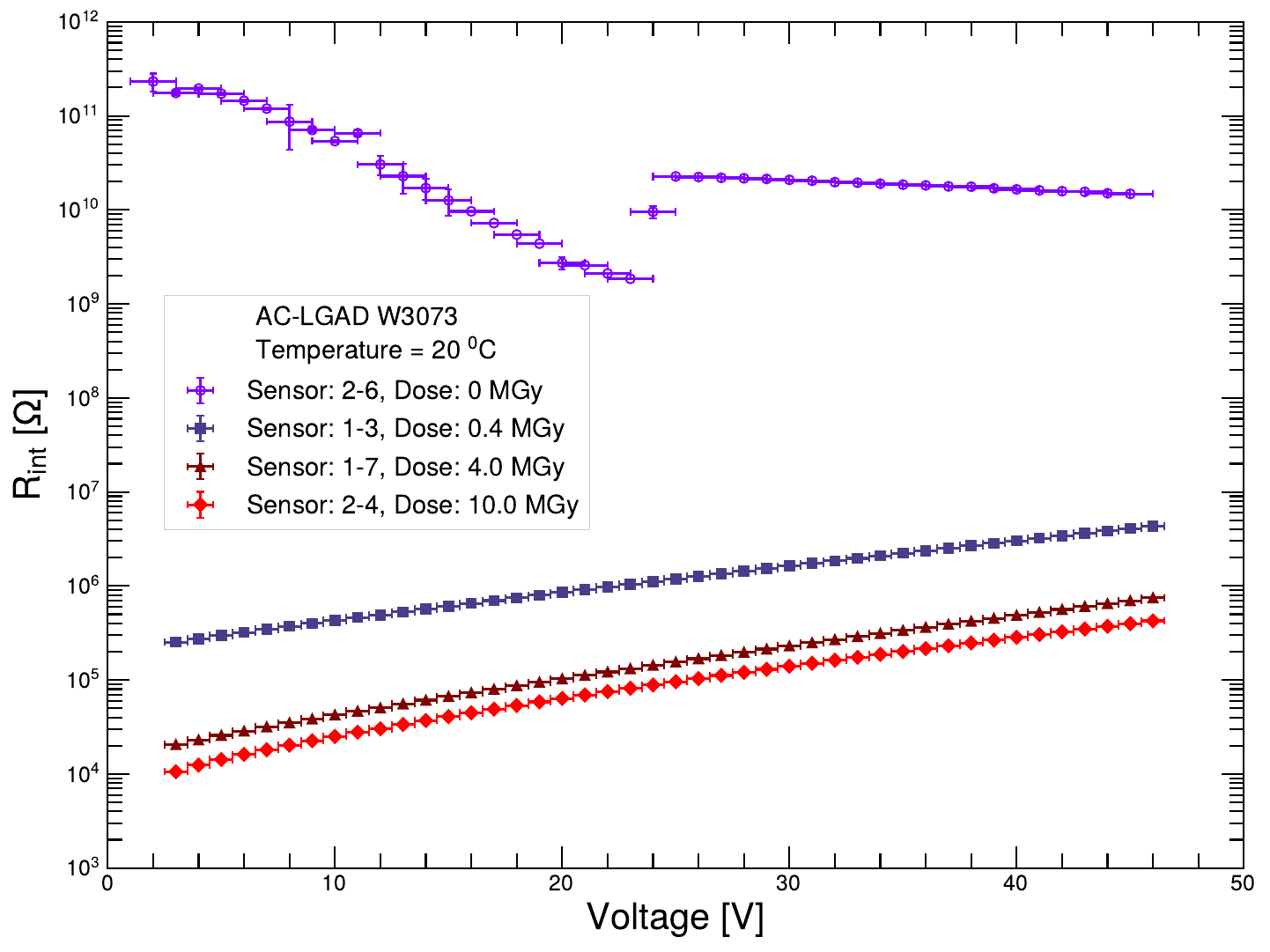}
\caption{Resistance from the guard ring to the DC pad, for strip AC-LGADs, as a function of bias voltage for several gamma dose values. The data were collected at approximately $20^\circ$C.
\label{fig:resistance_3073}}
\end{figure}

The data from the inter-pad resistance measurement are shown in Figure~\ref{fig:resistance_3073}.
In the unirradiated case, a dip in resistance occurs at the onset of full depletion.
In the depleted devices, the inter-pad resistance drops with increasing dose by about 4 orders of magnitude (from $10^{10}$ to below $10^6$ ohm), stabilizing above about 4~MGy.  

In the unirradiated case and below full depletion, the oxide charge density is relatively low and the resistance at low bias is 
dominated by the undepleted substrate and the resistive $n^{++}$ layer. Above full depletion, the substrate effects stabilize, and the resistance is fairly constant.  After irradiation with gammas, studies~\cite{oxide1,oxide2} have shown that the oxide charge density increases and saturates around 0.1 to 1~MGy.  The resistance is then dominated by the oxide charge after gamma exposure. The gamma exposure, besides causing surface damage, also generates defects in the silicon substrate though at a much lower level than does proton irradiation. The bulk damage is observed in Figure~\ref{fig:resistance_3073} in the lowering of the resistance from 0.4 to 10.0~MGy exposure.

\section{Proton Irradiation Studies}
\label{sect:proton}

\begin{figure}[htbp]
    \centering
\includegraphics[width=.45\textwidth]{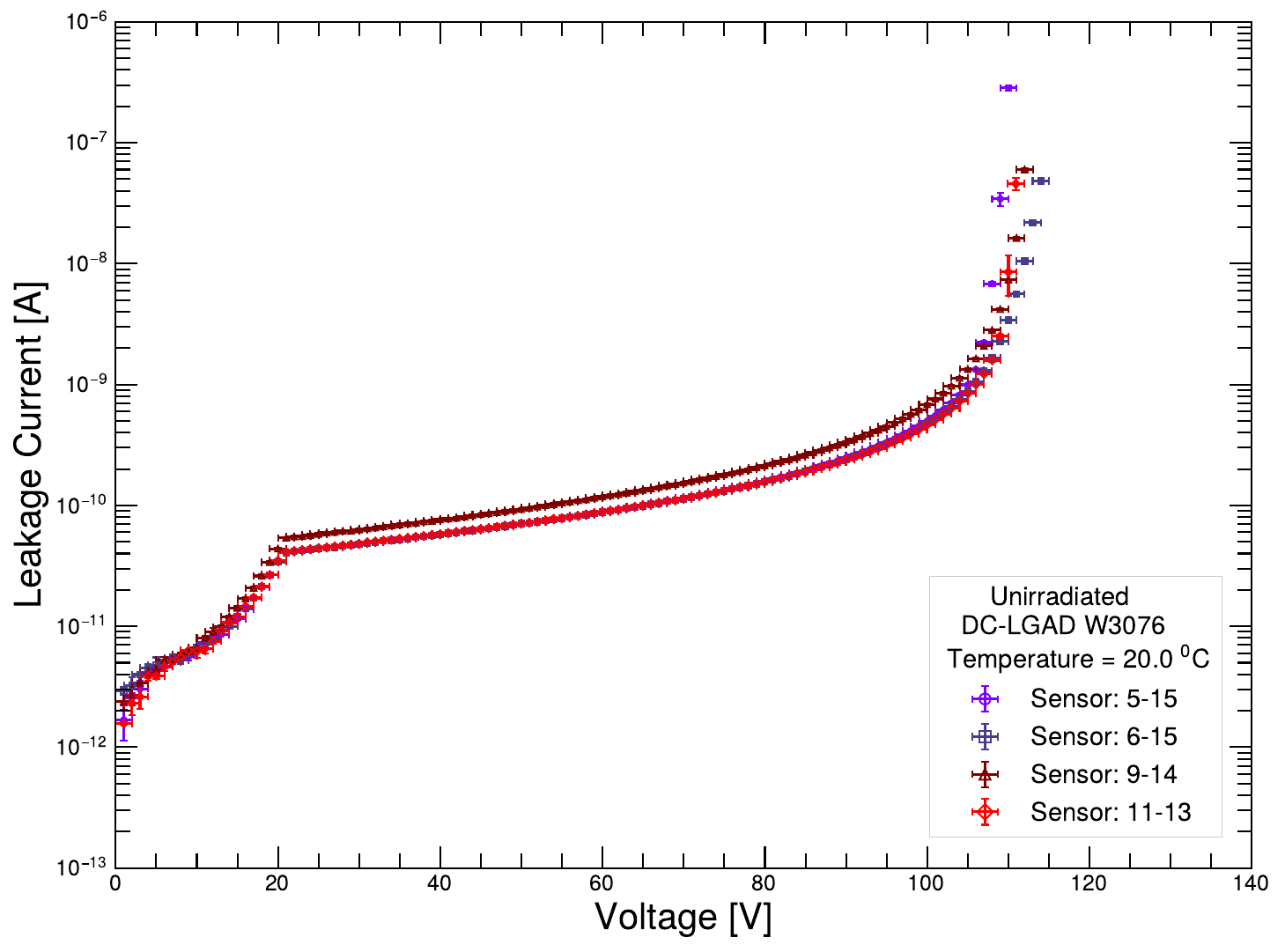}
\qquad
\includegraphics[width=.45\textwidth]{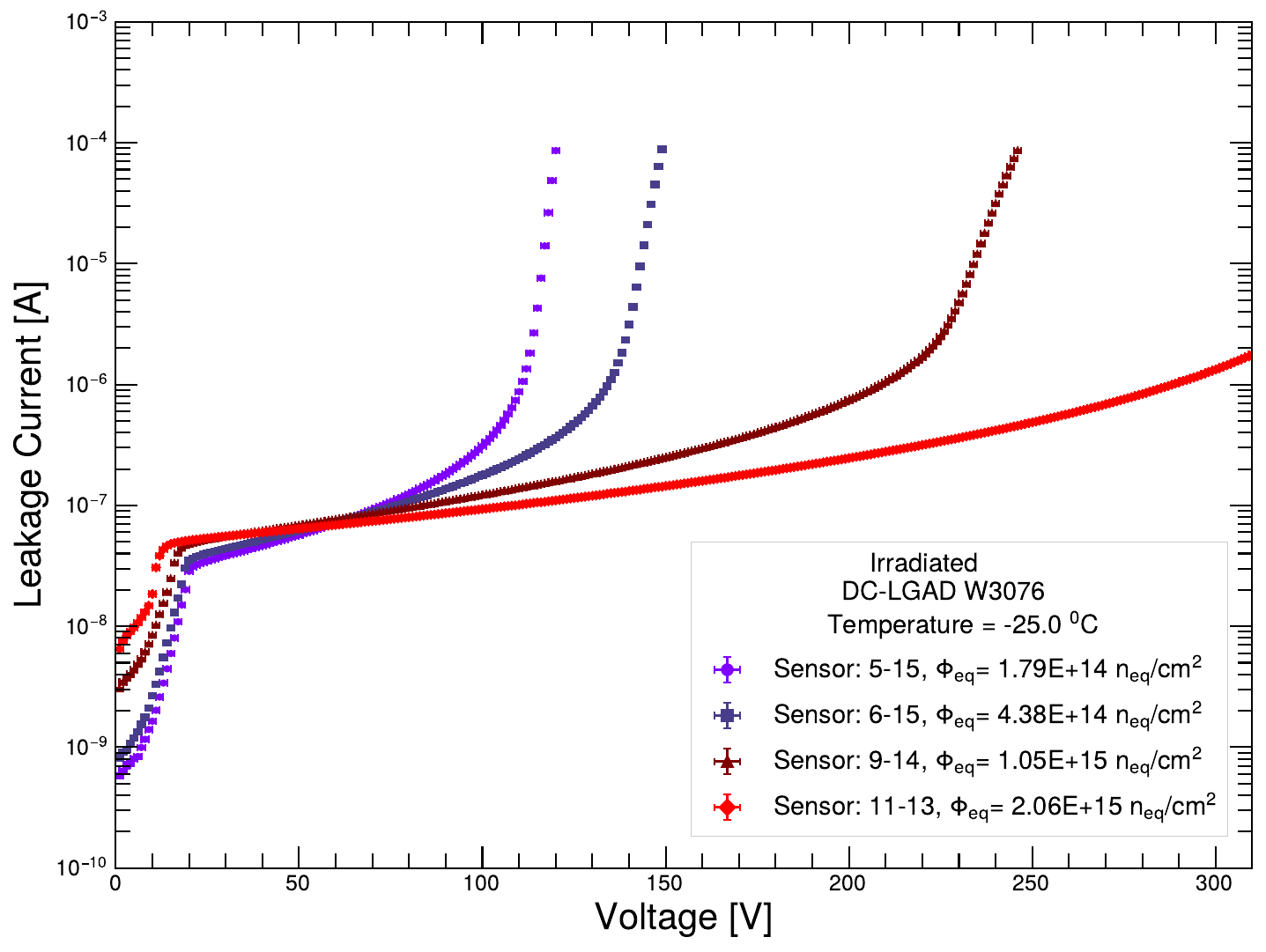}
\caption{Leakage current versus bias voltage of DC-LGAD pixel devices (W3076) unirradiated (left) and irradiated with 400-MeV protons (right). (The uncertainty bars on data points below the depletion voltage have been excluded for improved visualization.) The data on the left (right) were collected at temperatures in the range $20.0^\circ$C to $20.2^\circ$C ($-25.0^\circ$C to $-24.8^\circ$C), and are all scaled to $20^\circ$C ($-25^\circ$C) for ease of comparison.
\label{fig:W3076_IV}}
\end{figure}

\begin{figure}[htbp]
\centering
\includegraphics[width=.45\textwidth]{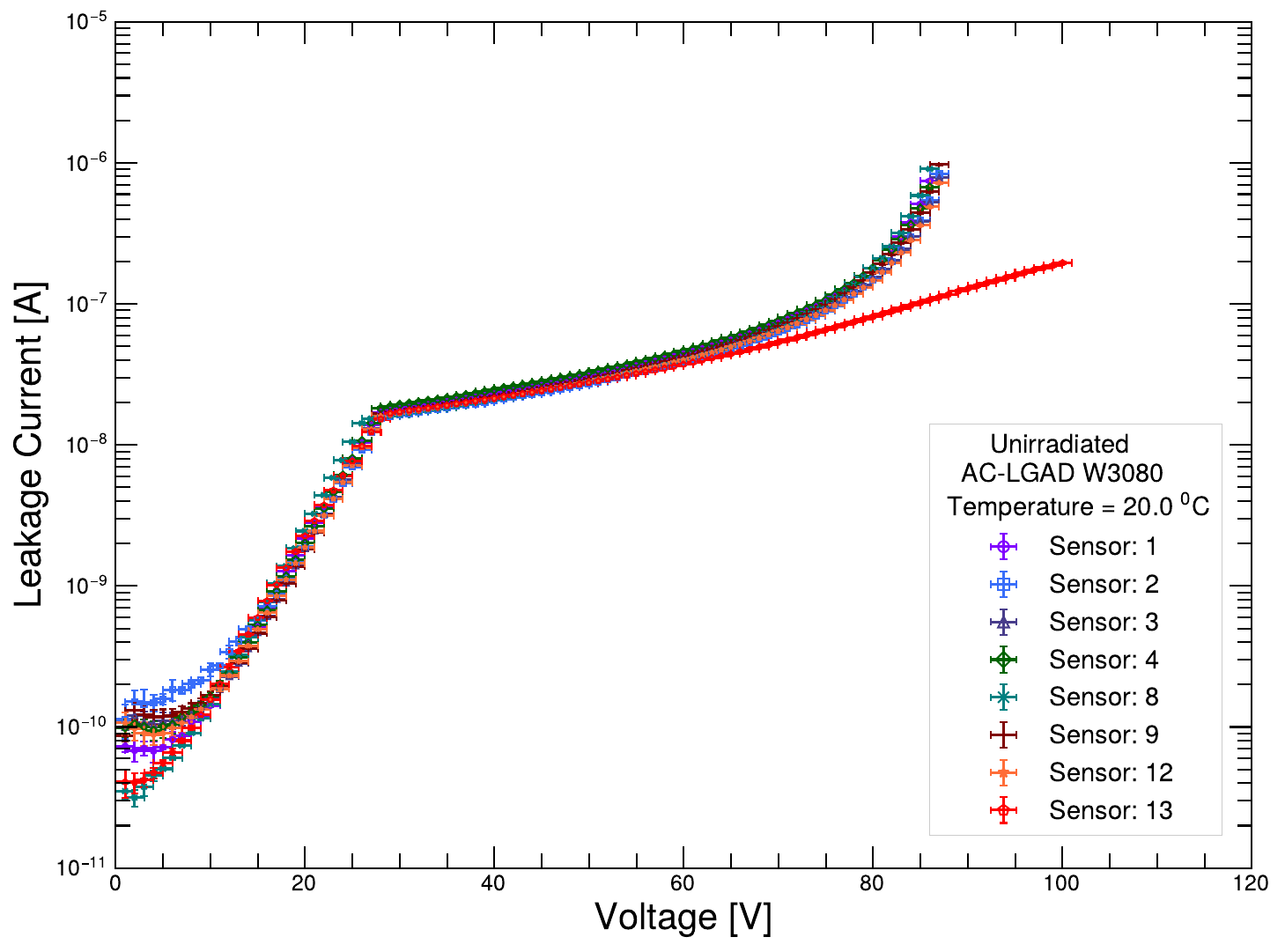}
\qquad
\includegraphics[width=.45\textwidth]{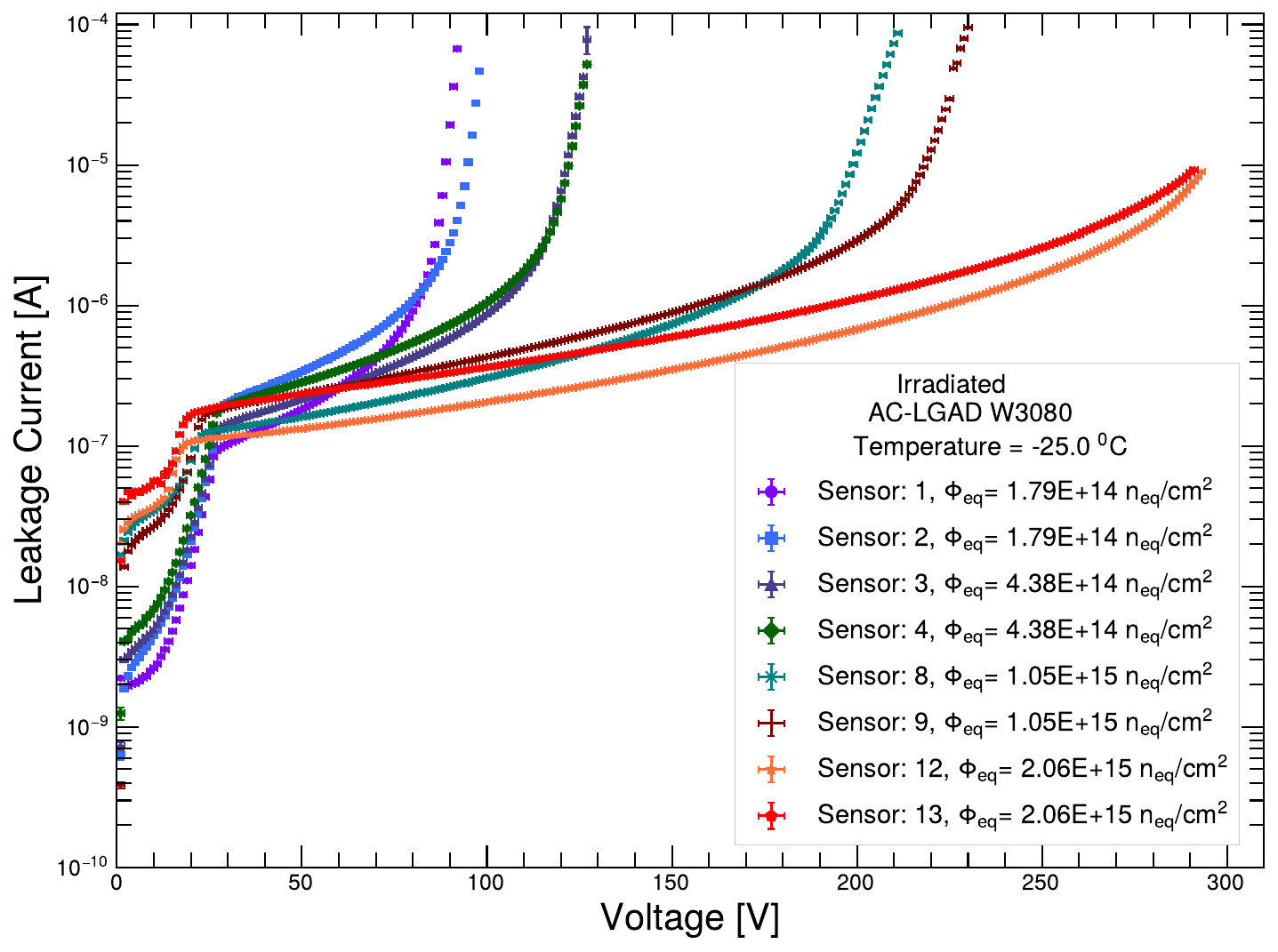}
\caption{Leakage current versus bias voltage of AC-LGAD strip devices (W3080) unirradiated (left) and irradiated with 400-MeV protons (right). (The uncertainty bars on data points below the depletion voltage have been excluded for improved visualization.)  The data on the left (right) were collected at temperatures in the range $19.9^\circ$C to $20.3^\circ$C ($-25.1^\circ$C to $-24.7^\circ$C) and are all scaled to $20^\circ$C ($-25^\circ$C) for ease of comparison.
\label{fig:W3080_IV}}
\end{figure}

Figure~\ref{fig:W3076_IV} shows the leakage current versus applied bias voltage of DC-LGADs exposed to five proton fluences from zero to $2.06 \times 10^{15} {\rm n}_{\rm eq}/{\rm cm}^2$.
Figure~\ref{fig:W3080_IV} shows the leakage current versus bias voltage of pixel AC-LGADs exposed to five proton fluences from zero to $2.06 \times 10^{15} {\rm n}_{\rm eq}/{\rm cm}^2$.

To characterize the effects of the irradiation on the leakage current of the samples shown in Figures~\ref{fig:W3076_IV} and \ref{fig:W3080_IV}, the data, which were collected at various temperatures, are all scaled to a reference temperature of $20^\circ$C with Equation~\ref{eq:temp_scaling_leakage}.  The scaled data are shown in Tables~\ref{tab:leakage_current_W3076} and \ref{tab:leakage_current_W3080}, respectively.  One sees that in the case of the AC-LGADs (Wafer 3080), application of 400-MeV protons in the range from 2 to $25 \times 10^{14} {\rm cm}^{-1}$ [1.79 to $20.6 \times 10^{14} {\rm n}_{\rm eq}/{\rm cm}^2$] increases the leakage current by a factor of approximately 765.  In the case of the DC-LGADs (Wafer 3076), application of 400-MeV protons in the range from 2 to $25 \times 10^{14} {\rm cm}^{-1}$ increases the leakage current by a factor of approximately 90,000.

\begin{table}[htbp]
\centering
\caption{Leakage current of DC-LGADs (W3076), scaled to $20^\circ$ C, measured at 50~V for various applied fluences.  $\mathcal R$ is the ratio $I_{\rm irradiated}/I_{\rm unirradiated}$. The uncertainties on these central values are discussed in Appendix~\ref{app:B}.  The uncertainty on $\mathcal R$ values is 3\%.}
\label{tab:leakage_current_W3076}
\begin{tabular}{cccccc}
\toprule
{Fluence} & {Fluence}& {Sensor} & {$I_{\rm unirradiated}$} & {$I_{\rm irradiated}$} & {$\mathcal R$} \\

[400-MeV p/cm$^2$]&[${\rm n}_{\rm eq}/{\rm cm}^2$]&number&[nA]&[$\mu$A]&\\

\midrule
2 $\times$ 10$^{14}$  & $1.79 \times 10^{14}$ &5-15  & 0.07 & 6.0 & 86000\\
5 $\times$ 10$^{14}$  & $4.38 \times 10^{14}$ &6-15  & 0.07 & 6.6 & 94000\\
13 $\times$ 10$^{14}$  & $1.05 \times 10^{15}$ &9-14  & 0.09 & 7.4 & 82000\\
25 $\times$ 10$^{14}$  & $2.06 \times 10^{15}$ &11-13 & 0.07 & 6.9 & 99000\\
\bottomrule
\end{tabular}
\end{table}

\begin{table}[htbp]
    \centering
    \caption{Leakage current of AC-LGADs (W3080), scaled to $20^\circ$ C, measured at 50~V for several applied fluences.  $\mathcal R$ is the ratio $I_{\rm irradiated}/I_{\rm unirradiated}$.  The uncertainties on these central values are discussed in Appendix~\ref{app:B}.  The uncertainty on $\mathcal R$ is 3\%.}
    \label{tab:leakage_current_W3080}
    \begin{tabular}{cccccc}
\toprule
{Fluence} & {Fluence}& {Sensor} & {$I_{\rm unirradiated}$} & {$I_{\rm irradiated}$} & {$\mathcal R$} \\
{[400-MeV p/cm$^2$]} &[${\rm n}_{\rm eq}/{\rm cm}^2$]&number & {[nA]} & {[$\mu$A]} & \\
\midrule
\multirow{2}{*}{$2 \times 10^{14}$} & \multirow{2}{*}{$1.79 \times  10^{14}$}& 1 & 31 & 19 & 619 \\
                                   && 2 & 27 & 26 & 952\\
\midrule
\multirow{2}{*}{$5 \times 10^{14}$} & \multirow{2}{*}{$4.38 \times 10^{14}$}& 3 & 29 & 23 & 783 \\
                                   && 4 & 33 & 30 & 921 \\
\midrule
\multirow{2}{*}{$13 \times 10^{14}$} & \multirow{2}{*}{$1.05 \times 10^{15}$}& 8 & 28 & 17 & 603 \\
                                    && 9 & 30 & 25 & 843 \\
\midrule
\multirow{2}{*}{$25 \times 10^{14}$} & \multirow{2}{*}{$2.06 \times 10^{15}$}& 12 & 28 & 14 & 496 \\
                                     && 13 & 28 & 25 & 904 \\
\bottomrule
\end{tabular}
\end{table}

\begin{figure}[htbp]
\centering
\includegraphics[width=.45\textwidth]{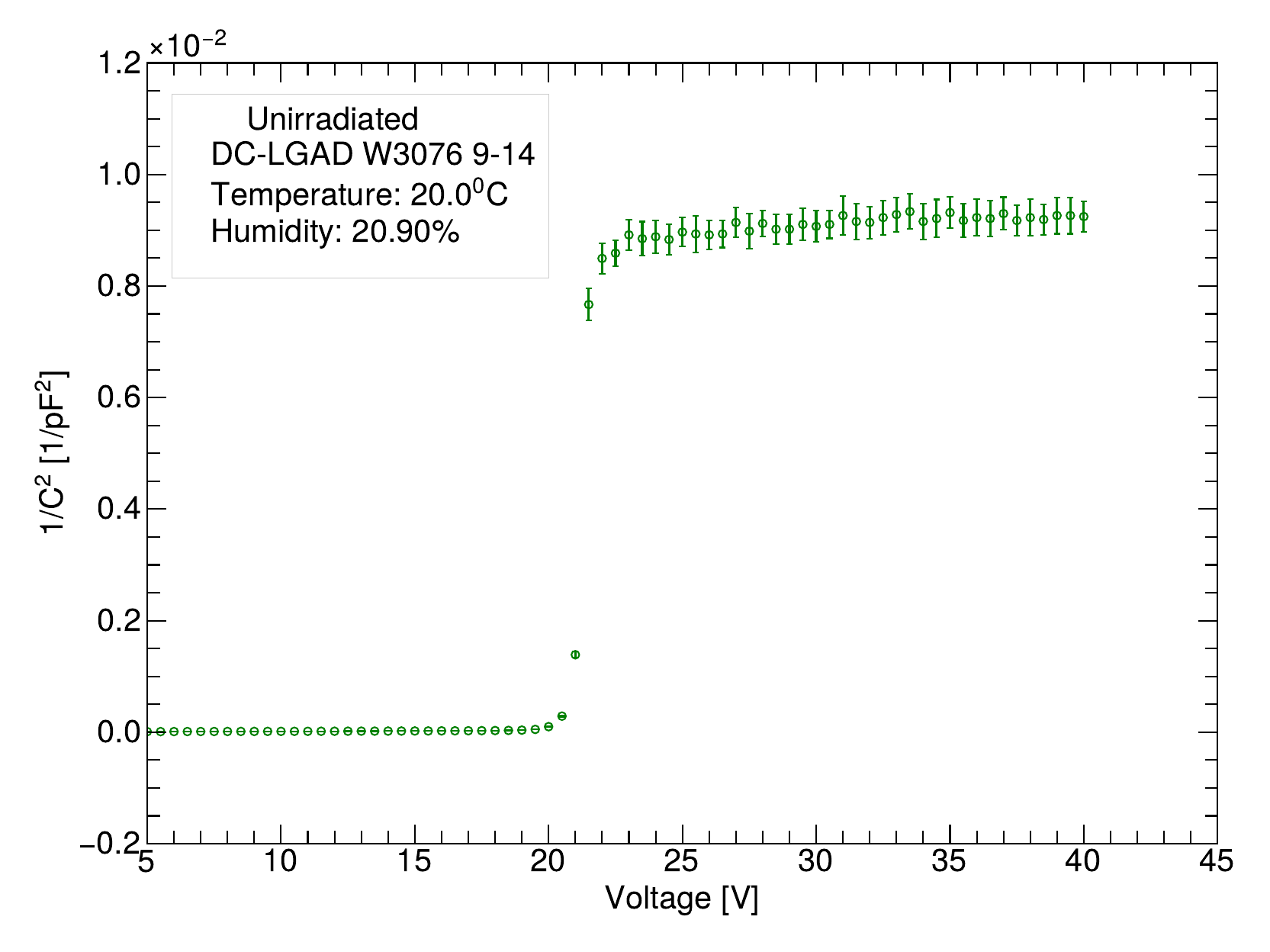}
\qquad
\includegraphics[width=.45\textwidth]{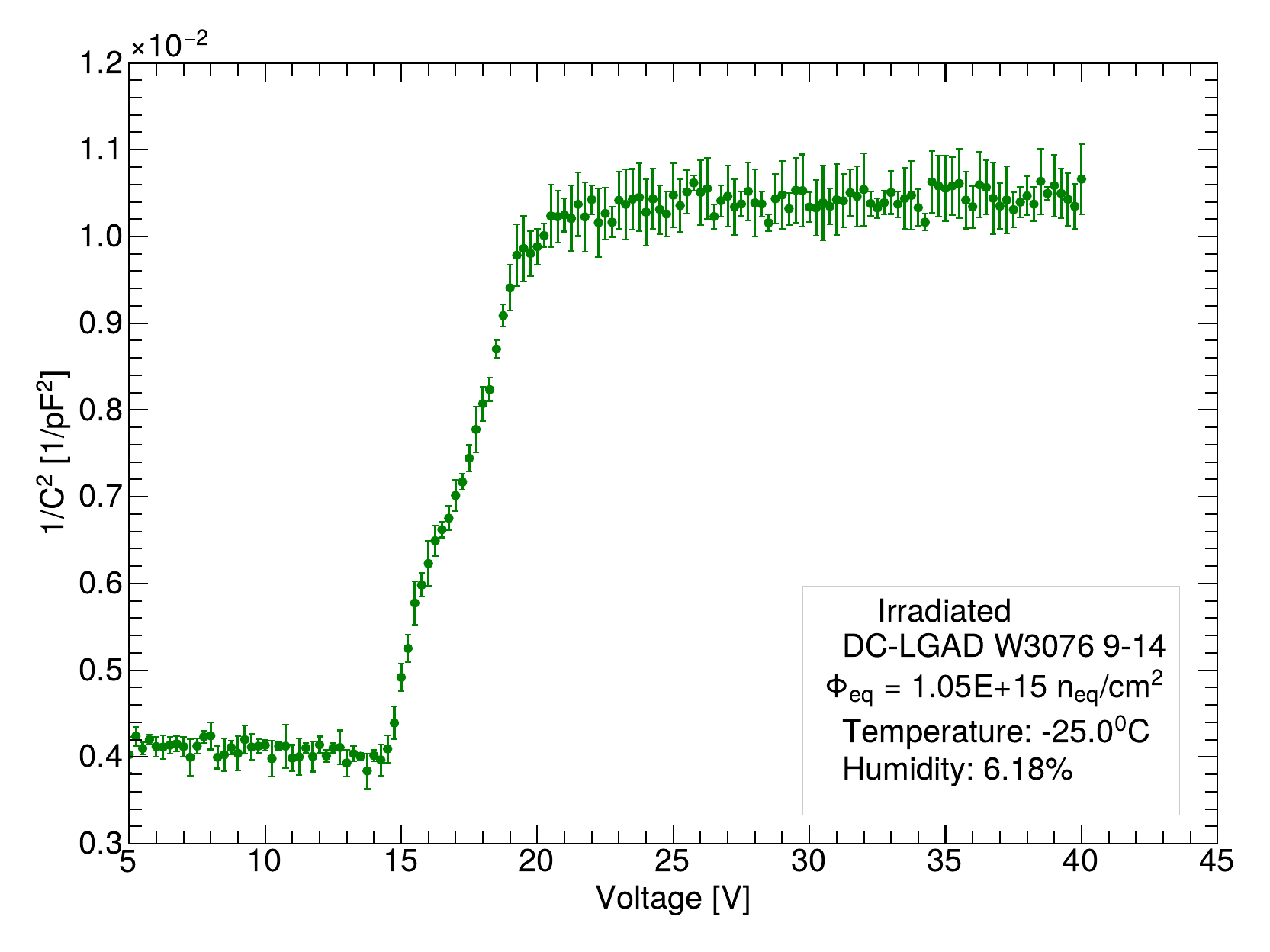}
\caption{Inverse capacitance (C$^{-2}$) versus bias voltage of DC-LGAD devices for two levels of 400-MeV proton exposure, recorded for signal frequency 1~kHz.  The unirradiated case is shown on the left, and the irradiated, on the right. The temperatures at which the data were collected are shown in the legends.
\label{fig:W3076_CV}}
\end{figure}

\begin{figure}[htbp]
\centering
\includegraphics[width=.45\textwidth]{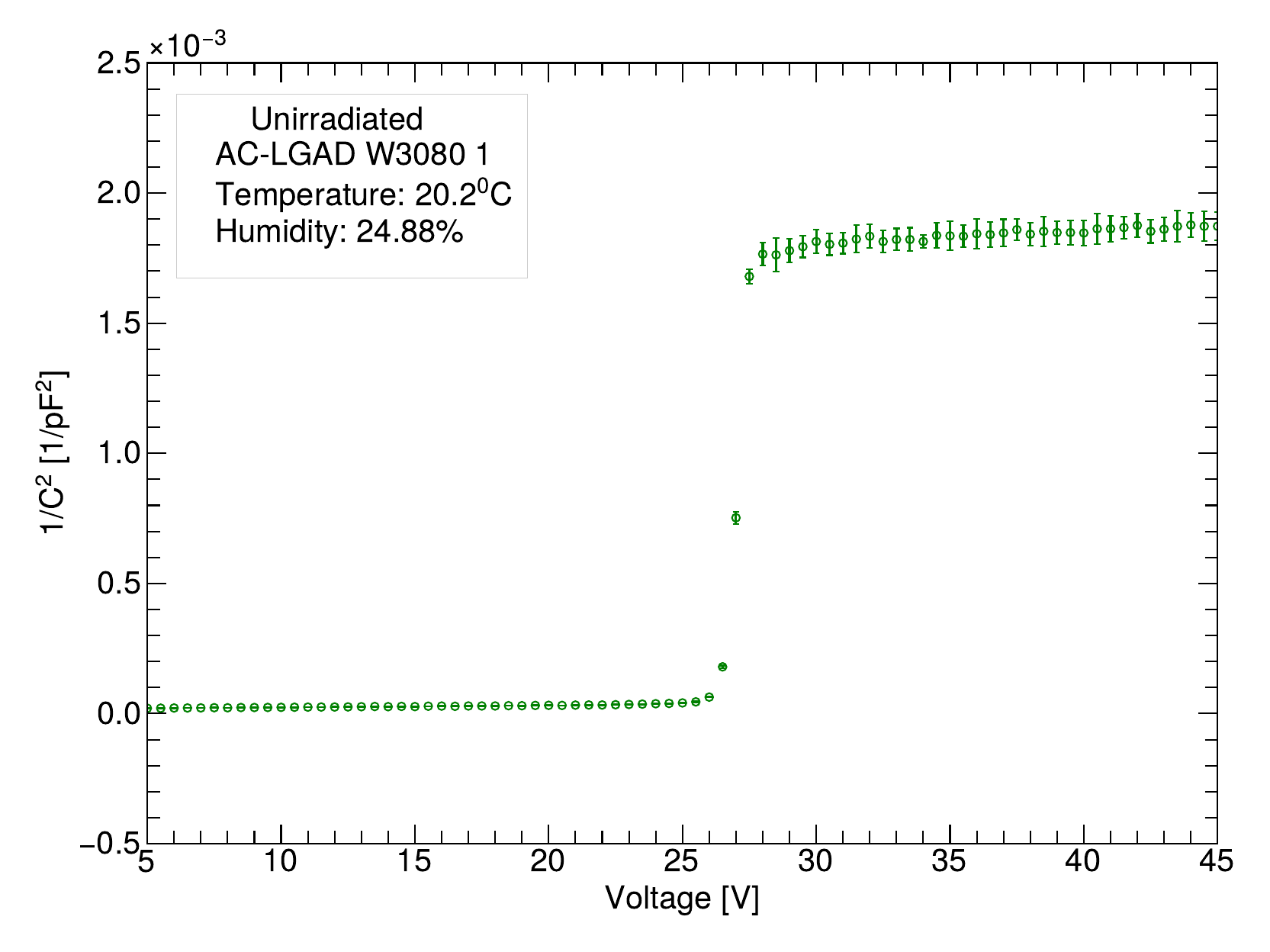}
\qquad
\includegraphics[width=.45\textwidth]{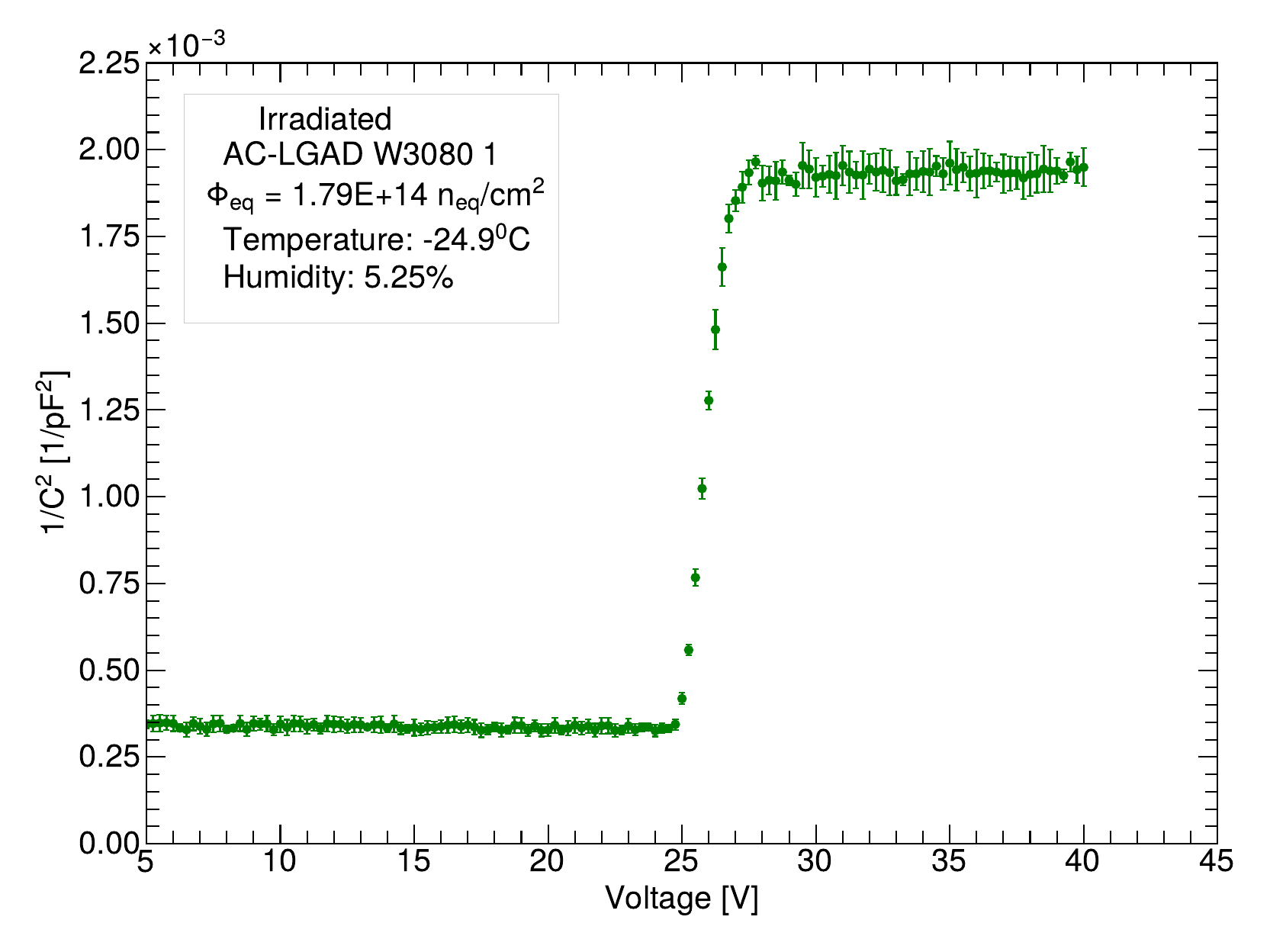}
\caption{Inverse capacitance (C$^{-2}$) versus bias voltage of AC-LGAD devices for two levels of 400-MeV proton exposure, recorded for signal frequency 1~kHz.  The unirradiated case is shown on the left, and the irradiated, on the right. The temperatures at which the data were collected are shown in the legends.
\label{fig:W3080_CV}}
\end{figure}

Figures~\ref{fig:W3076_CV} and \ref{fig:W3080_CV} show the CV characteristics for a typical proton-irradiated DC-LGAD and proton-irradiated AC-LGAD, respectively, recorded at applied signal frequency of 1~kHz.  (As with the gamma irradiated devices, a range of signal frequencies were studied.  Significant frequency-dependence in the CV trend was observed, and 1~kHz was judged to be the best condition for comparisons.)

The CV characteristics were used to extract depletion voltages from all samples for the full range of applied fluence.  
Tables~\ref{tab:Vgl_Vfd_proton_irr_W3076} and \ref{tab:Vgl_Vfd_proton_irr_W3080} compile the gain layer depletion voltage, full depletion voltage, and bulk depletion voltage before and after proton irradiation, for the DC-LGAD (Wafer 3076) and AC-LGAD (Wafer 3080), respectively.


\begin{table}[htbp]
\centering
\caption{Sensor voltages for DC-LGADs (W3076) after exposure to various 400-MeV proton fluences.  Sensor 11-13 sustained damage after the irradiation and could not subsequently be measured.}
\label{tab:Vgl_Vfd_proton_irr_W3076}
\small
\begin{tabular}{c|ccc|c|ccc}
\toprule
\multirow{2}{*}{\textbf{Sensor}} & \multicolumn{3}{c|}{\textbf{Unirradiated}} & \multirow{2}{*}{\begin{tabular}[c]{@{}c@{}}\textbf{Fluence} \\ $[{\rm n}_{\text{eq}}/\text{cm}^2]$\end{tabular}} & \multicolumn{3}{c}{\textbf{Irradiated}} \\ \cmidrule{2-4} \cmidrule{6-8} & $V_{\rm gl}$ [V] & $V_{\rm fd}$ [V] & $V_{\rm bulk}$ [V] & & $V_{\rm gl}$ [V] & $V_{\rm fd}$ [V] & $V_{\rm bulk}$ [V] \\ \midrule
5--15 & 20.3 $\pm$ 0.6 & 22.1 $\pm$ 0.6 & 1.8 $\pm$ 0.8 & 1.79E+14 & 19.1 $\pm$ 0.3 & 21.0 $\pm$ 0.3 & 1.9 $\pm$ 0.4 \\
6--15 & 20.3 $\pm$ 0.6 & 22.1 $\pm$ 0.6 & 1.8 $\pm$ 0.8 & 4.38E+14 & 17.8 $\pm$ 0.3 & 20.8 $\pm$ 0.3 & 3.0 $\pm$ 0.4 \\
9--14 & 20.3 $\pm$ 0.6 & 22.1 $\pm$ 0.6 & 1.8 $\pm$ 0.8 & 1.05E+15 & 14.2 $\pm$ 0.3 & 20.0 $\pm$ 0.3 & 5.8 $\pm$ 0.4 \\
11--13 & 20.2 $\pm$ 0.6 & 22.0 $\pm$ 0.6 & 1.8 $\pm$ 0.8 & 2.06E+15 & -- & -- & -- \\ \bottomrule
\end{tabular}
\end{table}


\begin{table}[htbp]
\centering
\caption{Sensor voltages for AC-LGADs (W3080) after exposure to various 400-MeV proton fluences.}
\label{tab:Vgl_Vfd_proton_irr_W3080}
\small
\begin{tabular}{c|ccc|c|ccc}
\toprule
\multirow{2}{*}{\textbf{Sensor}} & \multicolumn{3}{|c|}{\textbf{Unirradiated}} & \multirow{2}{*}{\begin{tabular}[c]{@{}c@{}}\textbf{Fluence} \\ $[{\rm n}_{\text{\rm eq}}/\text{cm}^2]$\end{tabular}} & \multicolumn{3}{c}{\textbf{Irradiated}} \\ \cmidrule{2-4} \cmidrule{6-8} & $V_{\rm gl}$ [V] & $V_{\rm fd}$ [V] & $V_{\rm bulk}$ [V] & & $V_{\rm gl}$ [V] & $V_{\rm fd}$ [V] & $V_{\rm bulk}$ [V] \\ \midrule
1 & 26.1 $\pm$ 0.5 & 27.9 $\pm$ 0.5 & 1.8 $\pm$ 0.8 & 1.79E+14 & 25.0 $\pm$ 0.3 & 26.8 $\pm$ 0.3 & 1.8 $\pm$ 0.4 \\
2 & 27.3 $\pm$ 0.6 & 29.4 $\pm$ 0.6 & 2.1 $\pm$ 0.9 & 1.79E+14 & 25.2 $\pm$ 0.3 & 26.6 $\pm$ 0.3 & 1.4 $\pm$ 0.4 \\
3 & 27.3 $\pm$ 0.6 & 29.4 $\pm$ 0.6 & 2.1 $\pm$ 0.9 & 4.38E+14 & 23.2 $\pm$ 0.3 & 26.6 $\pm$ 0.3 & 3.4 $\pm$ 0.4 \\
4 & 27.9 $\pm$ 0.5 & 29.1 $\pm$ 0.5 & 1.2 $\pm$ 0.7 & 4.38E+14 & 23.8 $\pm$ 0.5 & 26.6 $\pm$ 0.3 & 2.8 $\pm$ 0.7 \\
8 & 26.1 $\pm$ 0.5 & 27.9 $\pm$ 0.5 & 1.8 $\pm$ 0.8 & 1.05E+15 & 18.5 $\pm$ 0.6 & 24.6 $\pm$ 0.5 & 6.1 $\pm$ 0.8 \\
9 & 27.9 $\pm$ 0.5 & 29.1 $\pm$ 0.5 & 1.2 $\pm$ 0.7 & 1.05E+15 & 18.4 $\pm$ 0.3 & 24.4 $\pm$ 0.3 & 6.0 $\pm$ 0.4 \\
12 & 27.3 $\pm$ 0.5 & 29.3 $\pm$ 0.6 & 2.0 $\pm$ 0.8 & 2.06E+15 & 12.6 $\pm$ 0.7 & 19.8 $\pm$ 0.3 & 7.2 $\pm$ 0.8 \\
13 & 26.7 $\pm$ 0.6 & 28.7 $\pm$ 0.6 & 2.0 $\pm$ 0.8 & 2.06E+15 & 14.2 $\pm$ 0.3 & 20.3 $\pm$ 0.3 & 6.1 $\pm$ 0.4 \\ \bottomrule
\end{tabular}
\end{table}


Application of non-ionizing radiation in the few times $10^{14} {\rm n}_{\rm eq}/{\rm cm}^2$ range reduces depletion voltages in the DC-LGAD by about 1 to 3~V.  
In the case of the AC-LGAD, change of 1 to 3~V is observed in the depletion voltages up to about $10^{15} {\rm n}_{\rm eq}/{\rm cm}^2$; however from 1 to $2 \times 10^{15} {\rm n}_{\rm eq}/{\rm cm}^2$, the full depletion voltage and gain layer depletion voltage are shown to decrease by about 10~V. 

\section{Acceptor Removal Constants}
\label{sect:acceptor}

\begin{figure}[htbp]
\centering
\includegraphics[width=.45\textwidth]{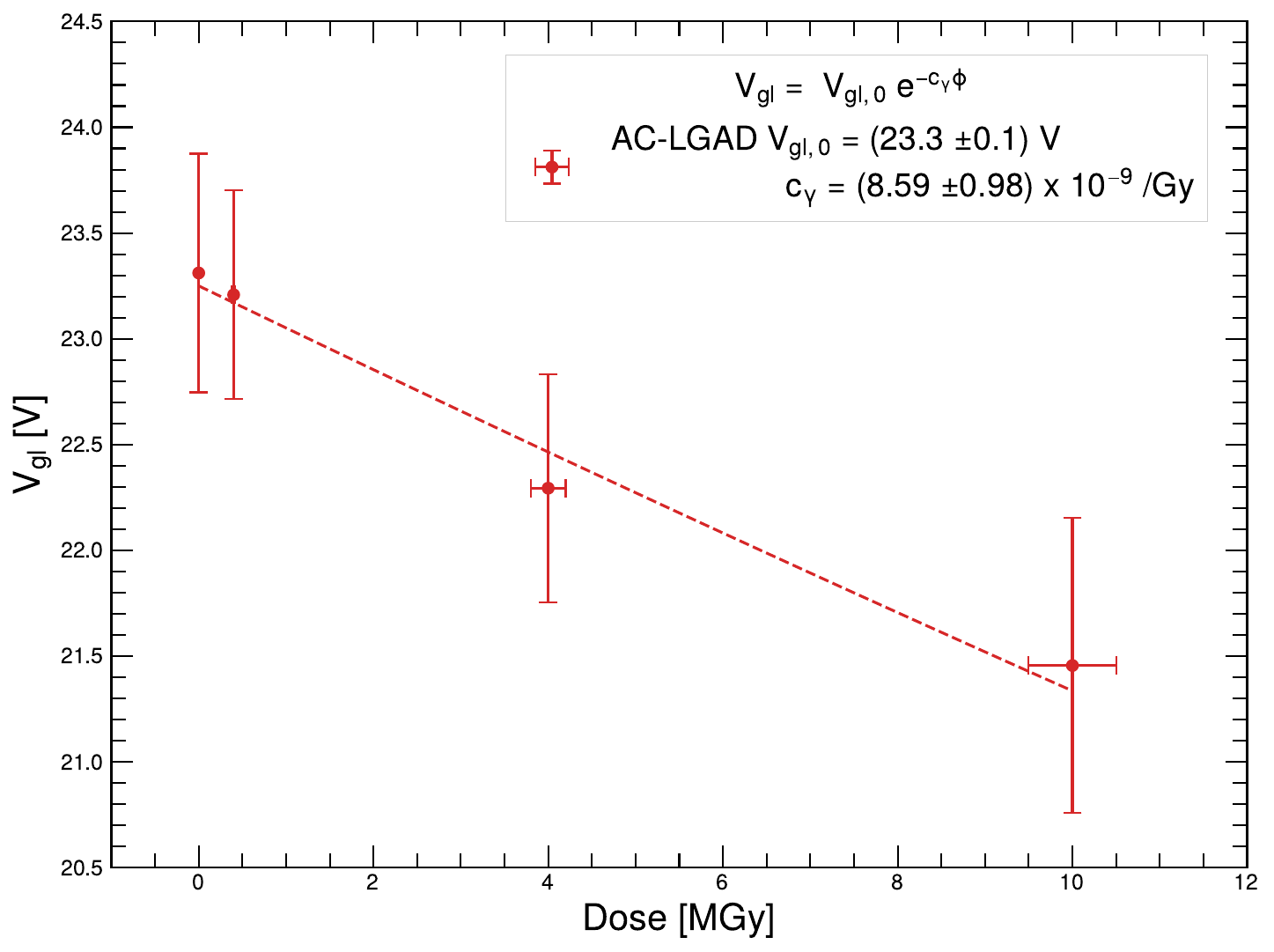}
\qquad
\includegraphics[width=.45\textwidth]{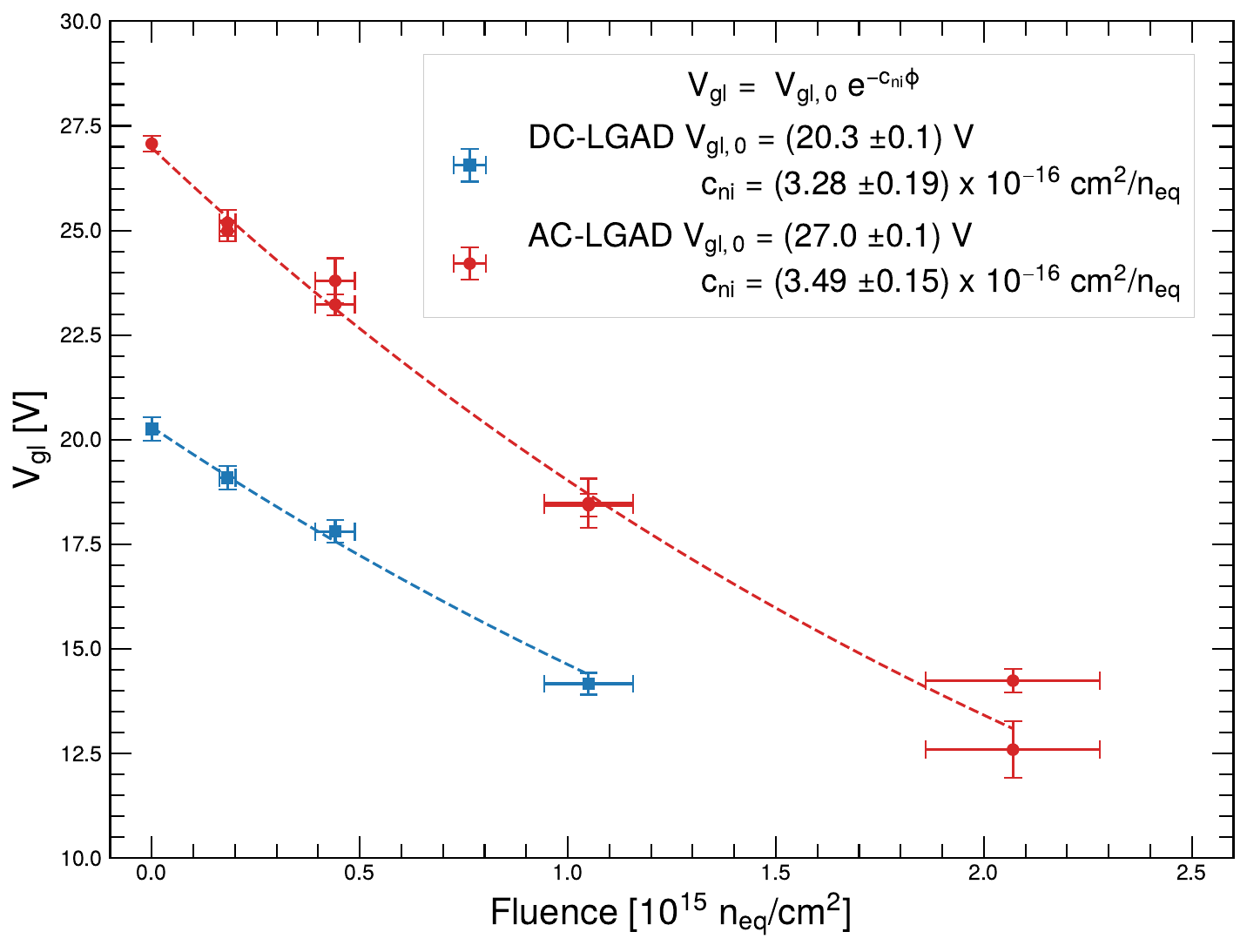}
\caption{(Left): Gain layer depletion voltage as a function of ionizing dose. The points represent
data obtained from fits of leakage current versus depletion voltage, and the curve represents the
function $V_{gl} = V_{gl,0} \cdot e^{-c_{\gamma}\phi}$. (Right): Gain layer depletion voltage as a function of non-ionizing
dose. The points represent data obtained from fits of leakage current versus depletion voltage,
and the curve represents the function $V_{gl} = V_{gl,0} \cdot e^{-c_{\rm ni}\phi}$.  The $V_{gl,0}$ is the depletion voltage prior to irradiation.  \label{fig:VglVsDose}}
\end{figure}

Acceptor removal refers to the process by which gain decreases as boron-substituted atoms deactivate due to radiation damage~\cite{acceptor1, acceptor2}.
For gamma-irradiated devices, the values of V$_{\rm gl}$ are shown in 
Figure~\ref{fig:VglVsDose} (left) as a function of dose.  The data in 
Figure~\ref{fig:VglVsDose} (left) are fitted to the function $V_{\rm gl}=V_{\rm gl,0}\cdot e^{-c_\gamma \phi}$ where $\phi$ is the total ionizing dose and c$_\gamma$  is the gamma-related acceptor removal constant.  The $V_{\rm gl,0}$ is the depletion voltage prior to irradiation.  For proton-irradiated devices, the values of V$_{\rm gl}$ are shown in 
Figure~\ref{fig:VglVsDose} (right) as a function of fluence.  The data in 
Figure~\ref{fig:VglVsDose}~(right) are fitted to the function $V_{\rm gl}=V_{\rm gl,0}\cdot e^{-c_{\rm ni} \phi}$ where c$_{\rm ni}$ is the acceptor removal constant due to non-ionizing radiation damage, and $\phi$ is the fluence.  Table~\ref{tab:ii} summarizes the acceptor removal constants that were obtained.  The uncertainties obtained above on V$_{\rm gl}$ and on the dose or fluence are propagated into the fitting function using orthogonal distance regression.
A typical total uncertainty on $V_{\rm gl}$ or $V_{\rm fd}$ is in the range 1\% - 6\% of the central value.  In Figure~\ref{fig:VglVsDose} one sees that the exponential decay of the ${\rm V}_{\rm gl}$ function describes the data well for both the DC- and AC-LGADs and for both ionizing and non-ionizing radiation.  The AC-LGADs have larger values of ${\rm V}_{\rm gl}$ than do the DC-LGADs. 

These acceptor removal constants may be compared with other measurements made under complementary conditions.  Those include values of $c_\gamma$ ranging from $1.47$ to $1.79 \times 10^{-8} {\rm Gy}^{-1}$, measured after exposure of wafers with four different gain layer options to 2.2~MGy~\cite{hoeferkamp}; and values of $c_{\rm ni}$ ranging from $3.1$ to $6.5 \times 10^{-16} {\rm cm}^2/{\rm n}_{\rm eq}$ for devices exposed to 400-MeV proton fluences up to $7.5 \times 10^{14} {\rm n}_{\rm eq}/{\rm cm}^2$~\cite{sorenson}.  The acceptor removal constants reported here are consistent with those in Refs.~\cite{hoeferkamp} and \cite{sorenson}.

\begin{table}[htbp]
\centering
\caption{Acceptor removal constants measured for three types of LGAD detectors.\label{tab:ii}}
\smallskip
\begin{tabular}{c|c|c}
\hline
AC-LGADs&DC-LGADs&AC-LGADs\\
wafer 3073&wafer 3076&wafer 3080\\
(c$_\gamma$)&(c$_{\rm ni}$)&(c$_{\rm ni}$)\\
\hline
$(8.59\pm0.98) \times 10^{-9} {\rm Gy}^{-1}$&$(3.28\pm0.19) \times 10^{-16} {\rm cm}^2/{\rm n}_{\rm eq}$&$(3.49\pm 0.15) \times 10^{-16} {\rm cm}^2/{\rm n}_{\rm eq}$\\

\hline
\end{tabular}
\end{table}

\begin{figure}[htbp]
\centering
\includegraphics[width=.45\textwidth]{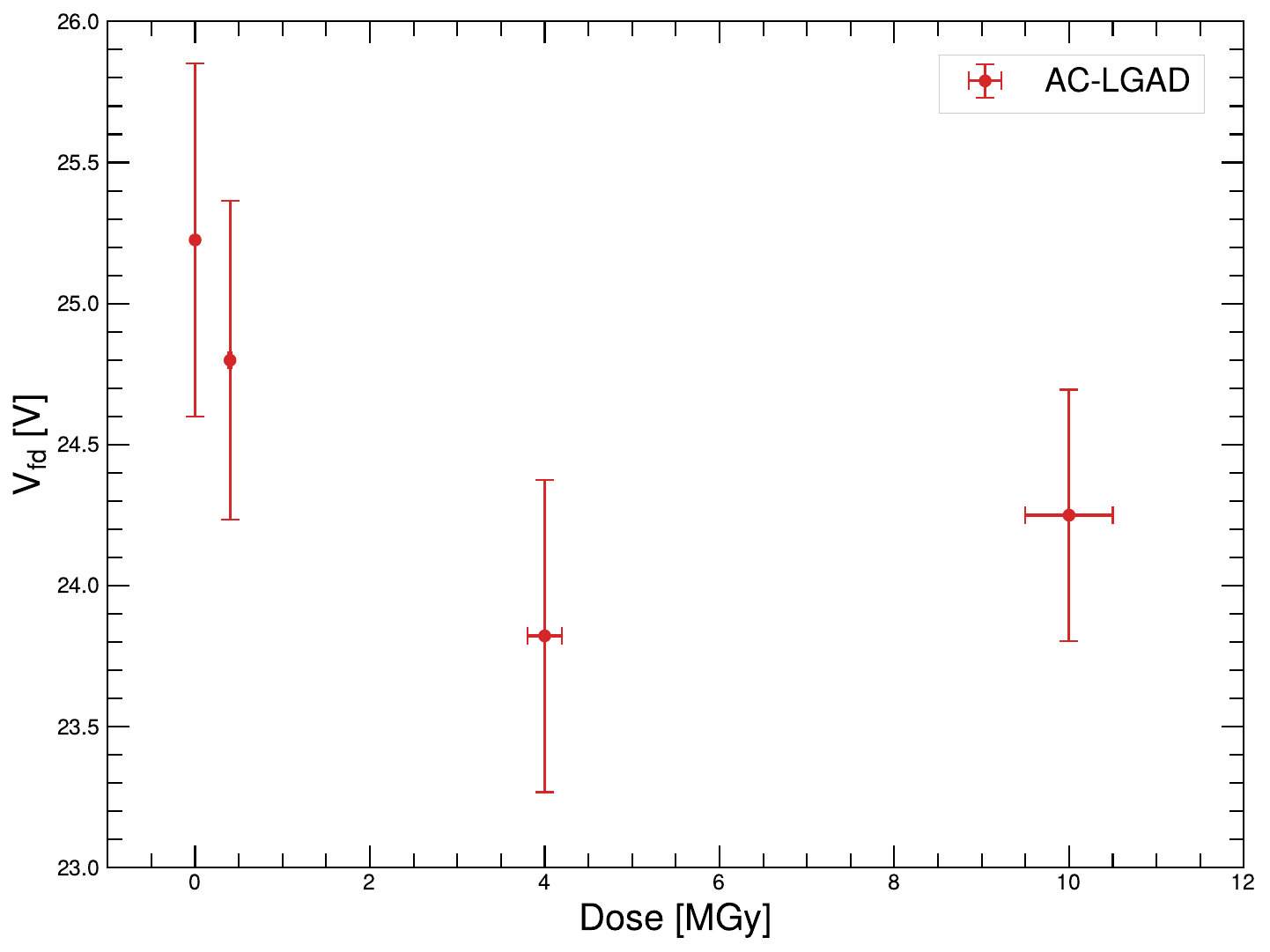}
\qquad
\includegraphics[width=.45\textwidth]{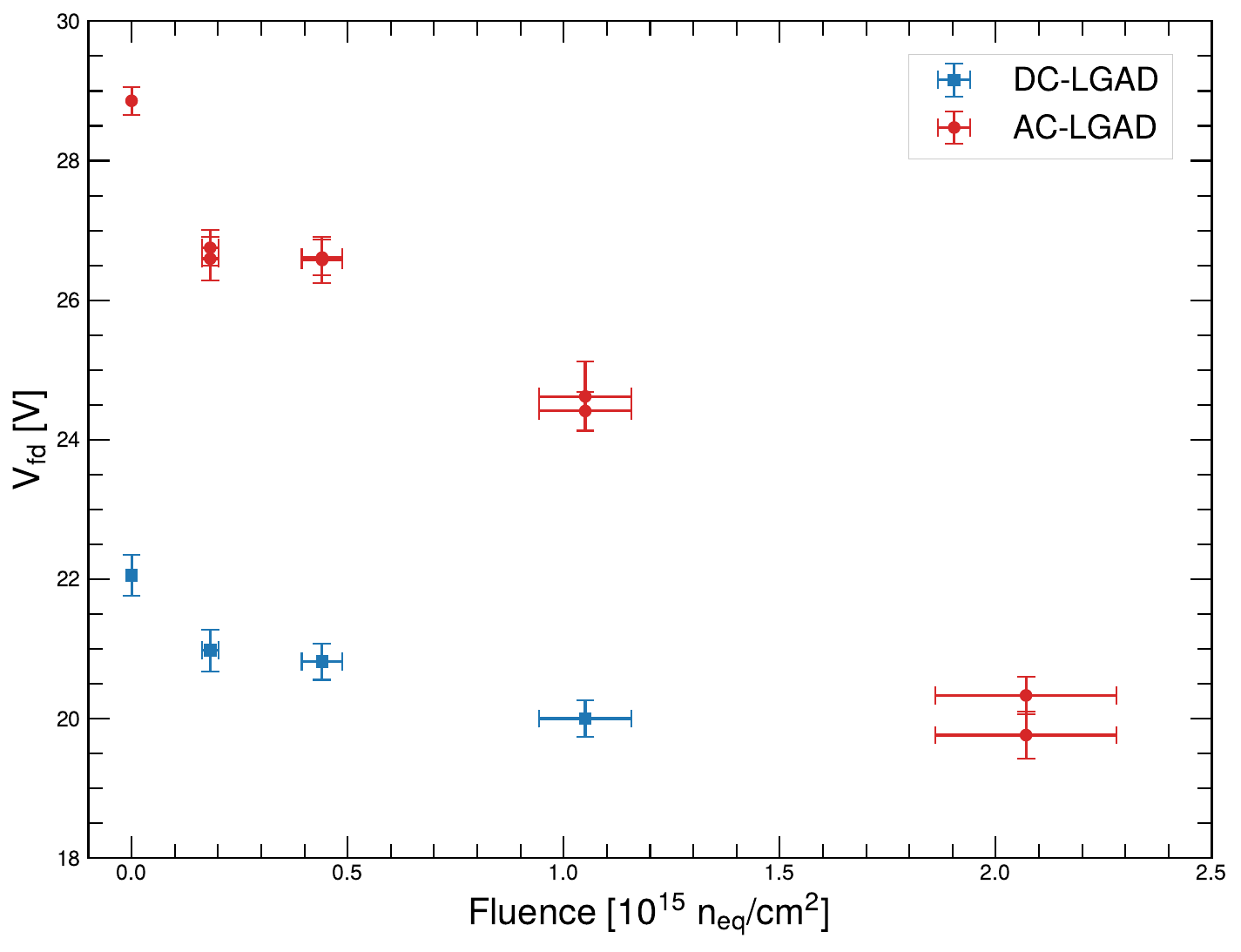}
\caption{(Left): Full depletion voltage as a function of ionizing dose of gamma irradiation. (Right): Full depletion voltage as a function of proton fluence. \label{fig:VfdVsDose}}
\end{figure}

Figure~\ref{fig:VfdVsDose} (left) shows the full depletion voltage of the AC-LGADs as a function of ionizing radiation dose.  Figure~\ref{fig:VfdVsDose} (right) shows the full depletion voltage of the DC-LGADs and AC-LGADs that were exposed to protons, as a function of proton fluence.  From Figure~\ref{fig:VfdVsDose} one sees that ${\rm V}_{\rm fd}$ decreases with increasing fluence in both DC- and AC-LGADs.  The AC-LGADs start with higher values of ${\rm V}_{\rm fd}$.

\section{Conclusions}
\label{sect:conclusions}

The IV and CV results are qualitatively consistent between pixels and strips after irradiation, and between DC- and AC-LGADs (neglecting effects due to variations in batch fabrication). Gamma irradiation up to 10 MGy appears to have a small effect on depletion voltages, possibly reducing them by 1~V.  Gamma irradiation introduces frequency dependence in the CV characteristic.

Irradiation to gamma doses of magnitude 40 (1000) MGy increases the leakage current in these devices at $20^\circ$C by a factor of about 3.5 (6.2).  The channel current rises from about 5~nA unirradiated to about 30 nA at 10 MGy. The breakdown voltage increases from about 120 V unirradiated to about 130 V at 10 MGy.

The resistance between the single DC contact and the guard ring was also measured as a function of gamma dose to evaluate the effects of surface damage due to ionizing radiation. In the unirradiated case, a dip in resistance occurs at the onset of full depletion.
In the depleted devices, the inter-pad resistance drops with increasing dose by about 4 orders of magnitude (from $10^{10}$ to below $10^6$ ohm), stabilizing above about 4~MGy. 

Irradiation with 400-MeV protons to fluences ranging from 2 to $25 \times 10^{14} {\rm cm}^{-2}$ [1.79 to $20.6 \times 10^{14} {\rm n}_{\rm eq}/{\rm cm}^2$] increases the leakage current at $20^\circ$C by a factor of about 90,000 in the DC-LGAD devices, and by a factor of about 765 in the AC-LGAD devices. Application of these levels of non-ionizing radiation reduces depletion voltages in the DC-LGAD by about 1 to 3~V.  In the case of the AC-LGAD, a similar decrease of a few volts is observed in the depletion voltages up to about $10^{15} {\rm n}_{\rm eq}/{\rm cm}^2$; however from 1 to $2 \times 10^{15} {\rm n}_{\rm eq}/{\rm cm}^2$, the full depletion voltage and gain layer depletion voltage decrease by about 10~V.  

The acceptor removal constants associated with ionizing and non-ionizing exposure of the AC-LGADs have been found to be $(8.59 \pm 0.98) \times 10^{-9} {\rm Gy}^{-1}$ and $(3.49 \pm 0.15) \times 10^{-16} {\rm cm}^2/{\rm n}_{\rm eq}$.  The acceptor removal constant associated with non-ionizing exposure of the DC-LGAD is consistent at $(3.28 \pm 0.19) \times 10^{-16} {\rm cm}^2/{\rm n}_{\rm eq}$.

\appendix
\section{Method for Determining Depletion Voltages $V_{\rm gl}$ and $V_{\rm fd}$}
\label{app:depletionVoltage}

\begin{figure}[htbp]
\centering
\includegraphics[width=.9\textwidth]{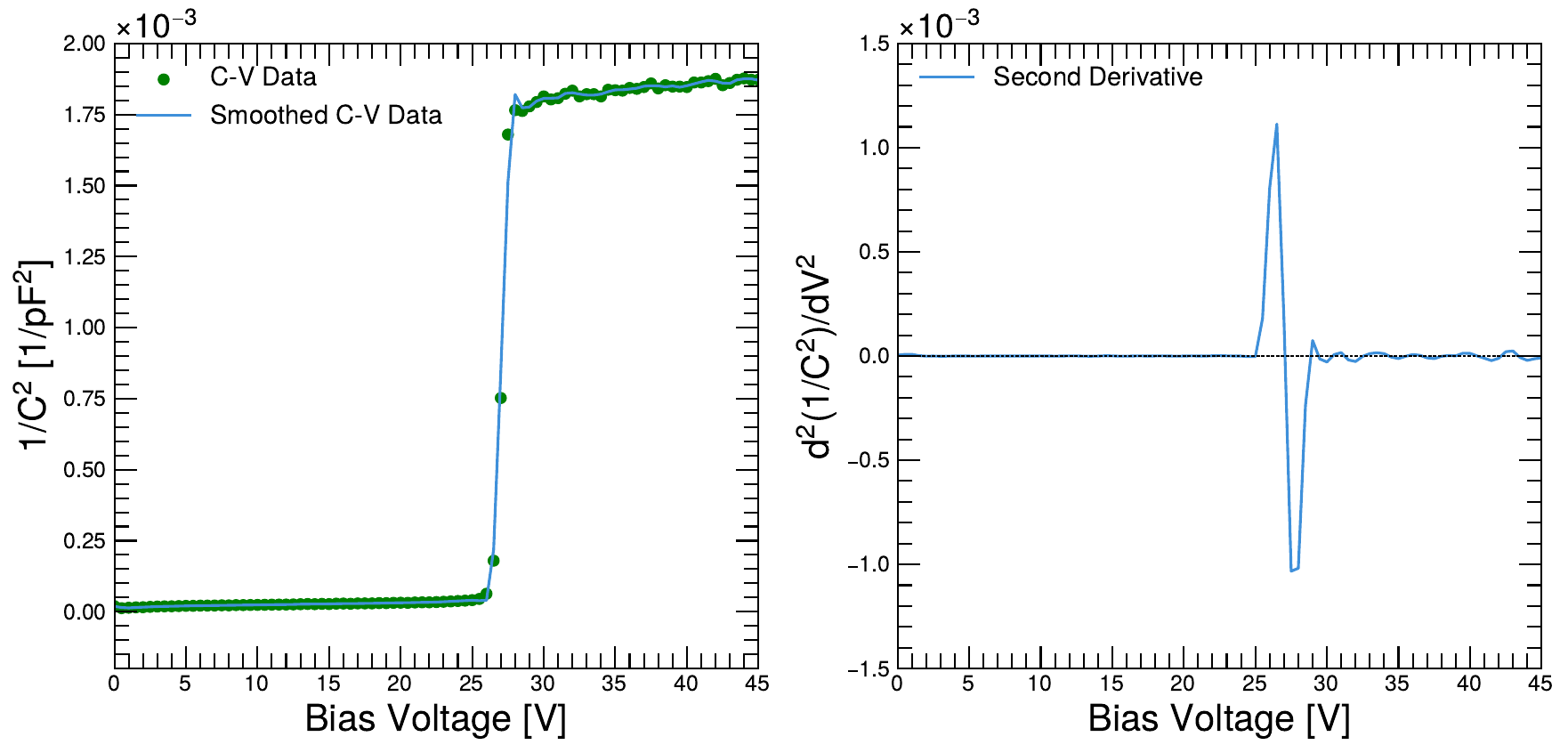}
\qquad
\includegraphics[width=.7\textwidth]{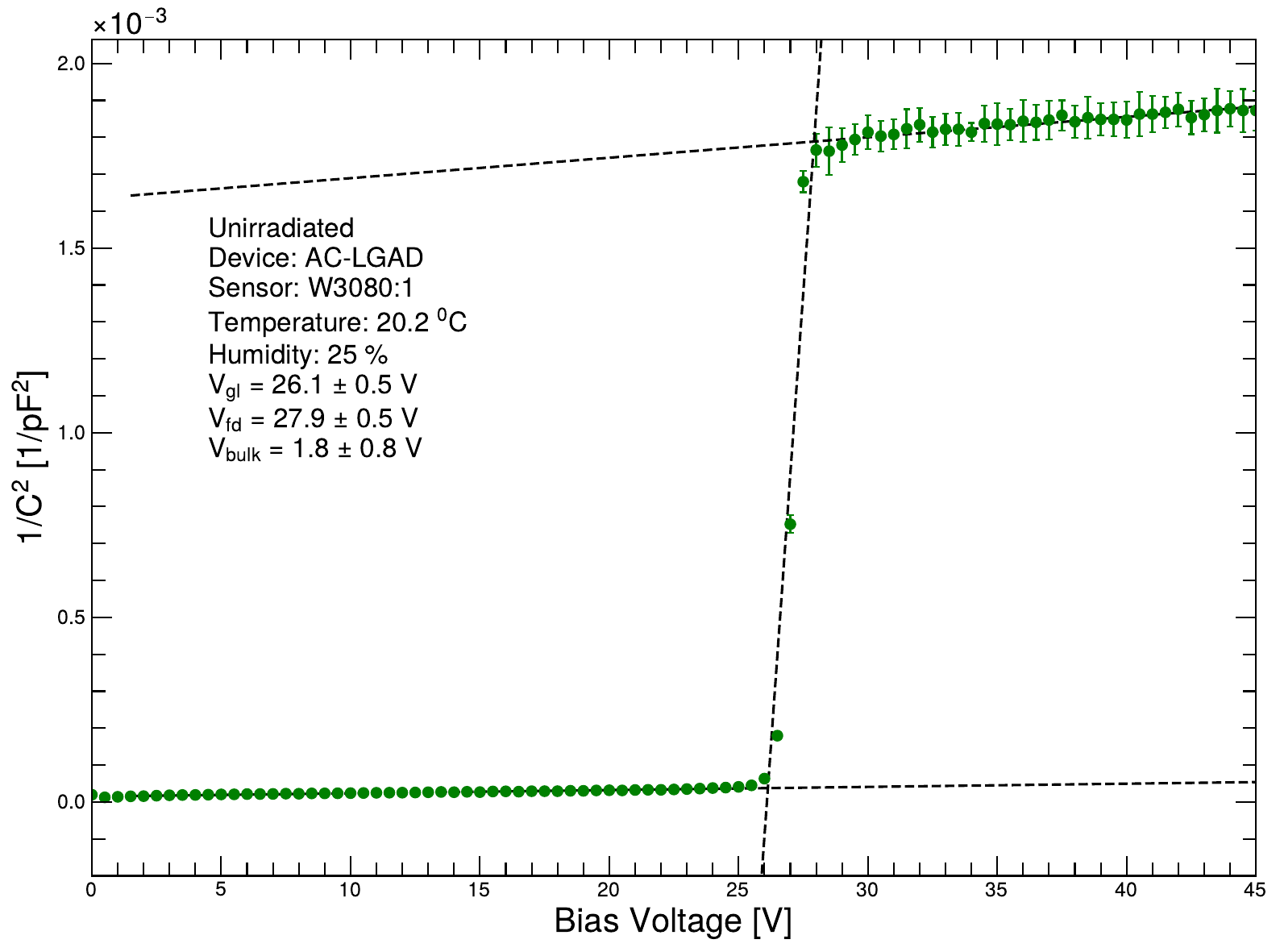}
\caption{(Upper left): The raw capacitance versus applied bias data are shown in green.  These are the same data as are also shown in Figure~\ref{fig:W3080_CV} (left). The smoothed function fitted to these data is shown in blue; (Upper right): The second derivative of the smoothed function representing the inverse squared capacitance with respect to applied bias is shown in blue.  The voltages $V_1$ an $V_2$ at which the upward-going and downward-going spikes occur, respectively, are recorded.  Those voltages define the limits of the data to which the three fitted lines are applied; (Bottom): Data selected from the smoothed function are used to fit three lines in the three distinct regions separated by $V_1$ and $V_2$. \label{fig:deplVolEst}}
\end{figure}

Figure~\ref{fig:deplVolEst} illustrates aspects of the procedure described in Sect.~\ref{sect:gamma} by which ${\rm V}_{\rm gl}$ and ${\rm V}_{\rm fd}$ are determined.  

\clearpage
\section{Systematic Uncertainties on Leakage Current, Capacitance, and Depletion Voltages}
\label{app:B}

The errors on the IV (CV) measurements include statistical and systematic uncertainties.  Each IV data point is the average of $N_I = 3$ to 5 measurements, and each CV data point is the average of $N_C = 3$ measurements.  The standard deviation $\sigma_I(\sigma_C)$ for IV (CV) measurements is found to be less than 1\% (less than 3\%).  Systematic uncertainties include uncertainties associated with the size of measurement steps: 1 V (0.5 V) for unirradiated devices, and 1 V (0.25 V) for irradiated devices.  The uncertainty on temperature is $\pm 0.5^\circ$C; this leads to a $\pm 1.82$\% uncertainty on leakage current~\cite{hoeferkamp} 
and negligible uncertainty contribution to capacitance.  
The instruments used to obtain the measurements contribute systematic uncertainties as well.  These are, for the Keithley 6487~\cite{manual} 
used to measure pad currents, 0.15\% for measurements on the 2~$\mu$A range, and 0.1\% for measurements on the 20~$\mu$A range, and for the HP4284A used to measure capacitance, $\pm 0.34$\%.

Thus the uncertainty on any measured leakage current data point is given by 

$$\Delta I_{\rm leakage}=\sqrt{{\sigma_I^2}+(0.0182\cdot I)^2+(0.0015\cdot I)^2}.$$

The uncertainty on any measured capacitance value is given by

$$\Delta C=\sqrt{{\sigma_C^2}+(0.0034\cdot C)^2}.$$

The total uncertainty shown on values of V$_{\rm gl}$ and V$_{\rm fd}$ is given by the quadrature sum of the statistical and systematic uncertainties.  
The statistical uncertainties on V$_{\text{gl}}$ and V$_{\text{fd}}$ are calculated using standard error propagation, accounting for the uncertainties and correlations of all parameters. 
Three fitted lines are described by equations $y=m_ix+c_i$, where the lines are numbered 1, 2, and 3 from left to right in the graph. The intersection points corresponding to V$_{\text{gl}}$ and V$_{\text{fd}}$ are given by:
$$V_{\text{gl}} = \frac{c_2 - c_1}{m_1 - m_2}$$
and
$$V_{\text{fd}} = \frac{c_2 - c_3}{m_3 - m_2}.$$
A linear least-squares fitting algorithm 
provides the uncertainties on the slope ($\Delta m_i$) and the intercept ($\Delta c_i$), as well as their correlation.
The statistical uncertainty on $V_{\text{bulk}} = V_{\text{fd}} - V_{\text{gl}}$ is calculated using 
$$\sigma_{V_{\text{bulk}}} = \sqrt{\sigma_{V_{\text{fd}}}^2 + \sigma_{V_{\text{gl}}}^2 - 2 \cdot \text{cov}(V_{\text{fd}}, V_{\text{gl}})}.$$
The covariance term arises because $V_{\text{gl}}$ and $V_{\text{fd}}$ are both derived from fits that share the middle line (Line 2).


Next, the middle line (Line 2) in the CV trend is fitted several times, each fit using a different number of data points, where the number ranges from the full set of points to a reduced set with up to ten points removed, while maintaining at least two points associated with that middle line.  After each fit, the values of V$_{\rm gl}$, V$_{\rm fd}$, and V$_{\rm bulk}$ are calculated.  
The preliminary systematic uncertainty on each depletion voltage is given by the largest difference among all values of each fitted parameter.  
This preliminary systematic uncertainty is then compared to the size of the voltage step used in the measurement, and the larger of the two values (preliminary systematic or voltage step) is taken as the final systematic uncertainty on that depletion voltage. 
Finally, the full uncertainty on each depletion voltage 
is taken as the quadrature sum of the statistical and systematic uncertainties.
The error on the leakage current ratio is obtained in the usual way from

$$
\Delta \mathcal R = \mathcal R \sqrt{\left(\frac{\Delta I_{\rm irradiated}}{I_{\rm irradiated}}\right)^2 + \left(\frac{\Delta I_{\rm unirradiated}}{I_{\rm unirradiated}}\right)^2},
$$
where $\mathcal R$ is the ratio $I_{\rm irradiated}/I_{\rm unirradiated}$.

\acknowledgments

This material is based upon work supported by the U.S.\ Department of Energy under grants DE-SC0012704, DE-SC443363, DE-SC426496 and DE-SC0020255.  This research used resources of the Center for Functional Nanomaterials, which is a U.S.\ DOE Office of Science facility, at Brookhaven National Laboratory under Contract No. DE-SC0012704.





\end{document}